\documentclass[twocolumn]{aastex63}

\usepackage{graphicx}
\usepackage{natbib}

\usepackage{booktabs}
\newcommand{\civ}{\ion{C}{4}}		% CIV
\newcommand{\oiii}{[\ion{O}{3}]}	% MgII
\newcommand{\lledd}{$L/L_{\rm Edd}$}	
\received{Dec.\ 5, 2020}
\revised{\today}
\accepted{}
\shorttitle{Probing Winds in Radio-Quiet Quasars}
\shortauthors{Richards et al.}

\graphicspath{./}
\begin{document}

% Title of the paper, and the short title which is used in the headers.
% Keep the title short and informative. (45 characters)
\title[Probing Winds in Radio-Quiet Quasars]{Probing the Wind Component of Radio Emission in Luminous High-Redshift Quasars}

\author[0000-0002-1061-1804]{Gordon T. Richards}
\affiliation{Department of Physics, Drexel University, 32 S.\ 32nd Street, Philadelphia, PA 19104, USA}

\author[0000-0002-9321-9559]{Trevor V. McCaffrey}
\affiliation{Department of Physics, Drexel University, 32 S.\ 32nd Street, Philadelphia, PA 19104, USA}

\author[0000-0001-9324-6787]{Amy Kimball}
\affiliation{National Radio Astronomy Observatory, 1003 Lopezville Rd, Socorro, NM 87801, USA}

\author[0000-0002-2091-1966]{Amy L. Rankine}
\affiliation{Institute of Astronomy, University of Cambridge, Madingley Road, Cambridge, CB3 0HA}

\author[0000-0002-3493-7737]{James H. Matthews}
\affiliation{Institute of Astronomy, University of Cambridge, Madingley Road, Cambridge, CB3 0HA}

\author[0000-0002-6528-1937]{Paul C. Hewett}
\affiliation{Institute of Astronomy, University of Cambridge, Madingley Road, Cambridge, CB3 0HA}

\author[0000-0001-8125-1669]{Angelica B. Rivera}
\affiliation{Department of Physics, Drexel University, 32 S.\ 32nd Street, Philadelphia, PA 19104, USA}



% Abstract of the paper (250 words)
\begin{abstract}

We discuss a novel probe of the contribution of wind-related shocks to the radio emission in otherwise radio-quiet quasars.  Given 1) the known non-linear correlation between UV and X-ray luminosity in quasars, 2) that such correlation leads to higher likelihood of radiation-line-driven winds in more luminous quasars, and 3) that luminous quasars are more abundant at high redshift, deep radio observations of high-redshift quasars are needed to probe potential contributions of accretion disk winds to radio emission from quasars.
To this end, we target a sample of 50 $z\simeq 1.65$ color-selected quasars that span the range of expected accretion disk wind properties as traced by broad \civ\ emission.  3-GHz observations with the Very Large Array to an rms of $\approx10\mu$Jy beam$^{-1}$ probe to star formation rates of $\sim400\,M_{\Sun}\,{\rm yr}^{-1}$, leading to 22 detections.  Supplementing these pointed observations are survey data of 388 sources from the LOFAR Two-metre Sky Survey Data Release 1 that reach comparable depth (for a typical radio spectral index), where 123 sources are detected.  These combined observations reveal a radio detection fraction that is a non-linear function of \civ\ emission line properties and suggest that the data may require multiple origins of radio emission in radio-quiet quasars.
Specifically, in radio-quiet quasars we find evidence for radio emission from weak jets or coronae in quasars with low Eddingtion ratios, with either (or both) star formation and accretion disk winds playing an important role in luminous quasars and correlated with increasing Eddington ratio.  Additional pointed radio observations will be needed to fully establish the nature of radio emission in radio-quiet quasars.

\end{abstract}

%\keywords{quasars: general -- quasars: emission lines}
%GTR: No longer used.  Instead use UAT upon submission.
%Radio quiet quasars(1354)
%Radio loud quasars(1349)
%Radio continuum emission(1340)
%Optical observation(1169)
%Emission line galaxies(459)

%%%%%%%%%%%%%%%%% BODY OF PAPER %%%%%%%%%%%%%%%%%%

\section{Introduction}

%Origin of radio in quasars not fully known
Despite the fact that the first quasars to be discovered were strong radio sources \citep{Schmidt1963,Condon+2013}, the origin of radio emission from quasars (particularly radio-quiet [RQ] quasars) remains an unsolved question.  Part of the problem is that there are likely multiple origins of radio emission, including jets, star formation, accretion disk coronae, and shocks from winds \citep[e.g.,][]{Simpson2017,Kimball,Panessa+19}, yet investigations tend to make the implicit assumption that strong radio emission is due to jets, while weaker radio emission is dominated by star formation.
Another part of the problem is that the fraction of luminous quasars detected in the radio is quite small.  \citet{Kellerman1989} examined 114 quasars from the Palomar Bright Quasar Survey
\citep{PG}---the so-called Palomar-Green (PG) quasar
sample---finding a radio-loud (RL\footnote{Herein, having high-frequency radio flux density greater than ten times the optical flux density or $\log R > 1$.}) fraction of just 15--20\%.  However, it is perhaps underappreciated that the ratio of RL to RQ quasars changes as a function of both redshift and luminosity \citep{Jiang2007,Kratzer2015}.  At low luminosity, \citet{HU01} and \citet{HP01} argue that the majority of Seyfert (\citeyear{Seyfert})
galaxies are RL sources.  At higher luminosity, the
fraction of quasars from the Sloan Digital Sky Survey (SDSS; \citealt{SDSS}) that are even {\em detected} (let alone formally RL) at the $\sim$1
mJy beam$^{-1}$ depth of the Faint Images of the Radio Sky (FIRST; \citealt{Becker1995}) survey is only 4\% \citep{DR7Q2010}.  The fraction of
{\em those} that are extended radio sources is only $\sim$ 10\%
(i.e., $\sim$0.4\% of SDSS quasars; \citealt{Ivezic2002}).

%Possible origins
For the luminous quasars that show radio lobes, it is clear that the dominant source of the radio emission is synchrotron emission from jets \citep{UP95}, thus it is necessary to break quasars into dichotomous RL and RQ classes (or jetted and non-jetted, \citealt{Padovani17}).  
However, it is equally important to understand that RQ quasars are not a monolithic class of objects either in terms of optical/UV spectra \citep{Richards2011} or radio emission \citep{Panessa+19}, and should, at a minimum, be divided into 
hard-spectrum and soft-spectrum RQ quasars \citep{Kratzer2015}.  
When probing to fainter radio flux densities, the (generally unresolved) radio emission is expected to be dominated by star formation (SF) processes that are largely unrelated to the central engine, either in the form of radio from free-free emission in \ion{H}{2} regions produced by star formation or from synchrotron emission produced by relativistic electrons accelerated by SNe \citep{Kimball+2011,Condon+2013}.  Both AGN and SF sources of radio emission might be expected to depend on environment \citep[e.g.,][]{Hickox+09,RMR17}.  
Beyond star formation, it is also important to determine if some quasars really are radio silent, if there is evidence for weak ``frustrated jets'' \citep[e.g.,][]{Barvainis+1996,BB98,hr+16}, coronal emission \citep{LB08,RL16,LBB19}, or shocks from winds interacting with gas in the host galaxy \citep[e.g.,][]{Jiang+10,ZG14,LBB19}. 
%BK07 is brehm not shocks?

%Radiation line driven winds
Thus, when investigating the radio demographics of active galactic nuclei (AGNs) it is crucial to make clear what part of the AGN landscape is being probed (e.g., host-dominated AGNs vs.\ luminous quasars).  Moreover, it must be established if the samples being investigated are capable of probing all radio origins.  For example, given that optical luminosity is seen to correlate with the presence of accretion disk winds (via the well-known optical to X-ray correlation; \citealt{Proga2000}), low-redshift samples such as those investigated by \citet{Kimball+2011} are not as capable of testing for wind-based radio emission.  
%\citep{BZ77,Blandford1982,WC95}
%\citealt{Narayan2003,Sikora2013,Rusinek2017,SR19}) 
Our approach is designed to particularly address radiation-line-driven winds thought to be active in the parent sample of broad absorption line quasars (BALQSOs; e.g.,  \citealt{Rankine+2020}).  
Early work on BALQSOs noted the possibility of an anti-correlation between the presence of winds (as indicated by broad absorption lines, \citealt{Weymann1991}) and the presence of radio jets \citep{Stocke1992}.  While there are examples of strongly radio-emitting BALQSOs \citep{Gregg+00,Brotherton+02,DiPompeo+11,Morabito2019}, there nevertheless persist significant differences (in properties related to accretion disk winds) between the most RL quasars and those with the most extreme absorption troughs \citep[][Figure 17]{Richards2011}.  
%Becker2000

%Groundwork for CIV approach
To test for a wind contribution to radio emission, we take advantage of the ability to recognize quasars whose spectra appear quite similar, but that have very different physical properties (e.g., as characterized by the Eddington ratio, \lledd; see \citealt{Sulentic+07}).   These differences are correlated with the observed radio properties of quasars \citep{Kratzer2015} and with the presence of winds \citep{Rankine+2020}.  \citet{Richards2011} and \citet{Kratzer2015} used two broad-emission-line diagnostics (the equivalent width, EW, of \civ\ and the ``blueshift" of the peak of \civ\ relative to the laboratory wavelength) to show that, while RL sources are not uniformly distributed across this ``\civ\ parameter space", neither do they occupy a distinct part of parameter space apart from radio-quiet quasars (in contrast to \citealt{Boroson2002}).  In any parameter space defined by optical/UV quasar properties, RQ quasars vastly outnumber RL quasars; however, the differences between the optical spectral properties of RL and RQ quasars are subtle at best \citep{Kimball+11b,Schulze+17}, spanning nowhere near the full range of spectral properties that characterize quasar spectra \citep{Boroson2002,Richards2011}.  
The primary goal of this investigation is to test for different possible origins
of radio emission specifically in luminous, high-redshift RQ quasars---from jets to shocks to coronoae to star formation---using deep radio observations to levels of radio emission expected to be dominated by star formation (at the redshift of
the quasars).  We focus on exploring the changes in radio luminosity and radio detection fraction as a function of \civ-based wind diagnostics in order to include shocks from winds as a potential source of radio emission \citep[e.g.,][]{jco+10,LBB19}, specifically in RQ quasars.

This paper is organized as follows.  In Section~\ref{sec:targets} we describe the targets, with Section~\ref{sec:data} presenting the details of the radio observations. 
Section~\ref{sec:c4space} outlines the empirical framework for the prediction of trends presented in Section~\ref{sec:predict}.  Sections~\ref{sec:sfr} discusses the relationships between the observed radio emission, emission line properties, and star formation.  
Finally in Sections~\ref{sec:discussion} and \ref{sec:conclusions} we discuss these results in the context of different proposed origins of radio emission and present our conclusions and suggestions for future work.  The cosmology assumed throughout the paper is $H_0=70$ km\,s$^{-1}$\,Mpc$^{-1}$, $\Omega_m=0.3$, and $\Omega_\Lambda=0.7$. The convention adopted for radio spectral index, $\alpha$, is $f_{\nu} \propto \nu^{\alpha}$ and the ratio of radio to optical emission, $R$, was calculated as the ratio between the rest-frame 3-GHz and 2500-${\rm \AA}$ flux density.

%However, it is important to realize that even in the regions where radio sources are more common, the radio-quiet sources still completely dominate the sample (that is, the 4\% detection fraction isn't the result of 100\% in some small corner of parameter space and nearly 0\% everywhere else).  Instead the range is more modest.  In this context, \cite{Richards2011} further noted that those quasars that occupy similar parts of CIV parameter space are curiously indistinguishable in terms of their optical spectra---regardless of whether they are RL or radio quiet (RQ).  

\section{Targets}
\label{sec:targets}

%PCH e-mail: 26 January 2018
%"In case makes life easier I have taken the DR7 quasars z=1.65-2.20 and selected those with a reasonable spectrum S/N (>=10 per pixel over 1600-3000A) to give 8653 quasars."

We need a representative sample
that is capable of probing multiple origins of radio emission in RQ quasars, specifically
%that overcomes the biases of the PG quasar sample \citep[e.g.,][]{Jester+2005}, which still strongly influences much of what we know about quasars today, including their radio properties.
that can address the question of whether shocks from accretion disk winds can be contributors to radio emission in luminous quasars.  
Theoretical understanding of radiation-line-driven accretion disk winds \citep{Murray1995,Proga2000} requires an abundance of UV photons (responsible for driving the wind) and a dearth of X-ray photons (that limit line driving) for strong radiation line driving.  The empirical observation of an anti-correlation between the optical/UV and X-ray \citep[e.g.,][]{Steffen2006} means that our investigation requires high luminosity sources (in the optical/UV). The small volume available at low redshift translates into the need for high-redshift ($z\gtrsim 1$) sources.  Over roughly $1.65< z < 2.2$ the \civ\ emission line serves as a diagnostic indicating the presence of accretion disk winds \citep{Richards2011}.  

At the inception of this investigation, there was a dearth of deep radio observations (e.g., as compared to those from the Faint Images of the Radio Sky [FIRST; \citealt{Becker1995}] survey), covering large enough area to include large numbers of high-redshift quasars.  Thus we proposed and were awarded pointed observations with the Karl G.\ Jansky Very Large Array (VLA; \citealt{Perley+11}); see Sections~\ref{subsec:VLAExperiment}--\ref{sec:forced}.  Since that time, data from the LOw-Frequency ARray (LOFAR; \citealt{LOFAR}) has become public and will be included as well (Section~\ref{sec:LOFARdata}).  A complementary investigation based entirely on the first data release of the LOFAR Two-metre Sky Survey (LoTSS; \citealt{Shimwell2019})
is presented by \citet{Rankine+2021}.

The targets were defined starting with the SDSS-DR7 quasar sample \citep{DR7Q2010}.  We imposed a limit on spectral signal to noise of ${\rm S/N}\ge10$ per pixel over a window of 1600--3000\AA\ in the restframe, giving a sample of 8653 quasars.  To avoid biases inherent to the fact that some SDSS quasars were selected explicitly because of their radio emission \citep{Richards+2002a}, we included only 8403 SDSS quasars that were targeted by color selection (but not excluding quasars that were {\em also} radio selected, indeed two of the targets are known FIRST sources).  We prefer to concentrate on the SDSS-DR7 sample, rather than the SDSS-DR16 \citep{DR16} sample that is currently available as DR16 is 
based on less homogenous selection criteria.

In practice another important consideration is whether the quasars are part of the ``uniform" selection criteria (which implies $i_{\rm AB} <19.1$), see \citet{Richards+2002a} and \citet{rsf+06}.  While we have chosen not to include that constraint in our investigation, it could be important for future work that attempts to explore radio emission from winds over a larger range in optical luminosity (where sources may be fainter than the magnitude limit of the ``uniform" sample).
%however, we have not imposed that constraint.
%In short, the target selection was changing in the early days of the SDSS survey and ``uniform" objects were selected using the final algorithm as described in \citet{Richards+2002a} and \citet{rsf+06}.  It is possible to use objects that are not included in the uniform sample for our analysis (especially if limiting to color-selection which helps constrain the uniformity), but proper statistical analysis should be limited to the uniform sample.  
The spectral reconstruction approach we have taken here 
(see \S~\ref{sec:c4space}) means that we have also not excluded quasars where broad absorption features \citep{Weymann1991} corrupt the profile of the emission line when using more common analysis techniques.

%[GTR: N.B. That our carefully selected sample of quasars would have required a survey of at least 50 times the area of Seymour+04 given that there was only 1 quasar with 1.5<z<1.8 among the 449 radio soruces.]

\section{Radio Data}
\label{sec:data}

\subsection{VLA Experimental Design}
\label{subsec:VLAExperiment}

%/Users/gtr/Dropbox/research/vla2020/vlatest2.ipynb

For the sake of identifying a suitably sized pilot sample for deep observations with the VLA, we 
further limited the VLA targets to $1.645\le z \le 1.6519$, which is the smallest range of redshift at the lower end of the distribution that yielded the desired pilot sample of 50 quasars.  The small redshift window has the added benefit of minimizing any redshift evolution in the sample. %[GTR: HW10 redshifts?  or SDSS?  Though it doesn't matter much here.]  
These targets span a range of bolometric luminosities\footnote{Using Equation~9 from \citet{Krawczyk+13}.} of $\log L_{\rm Bol}/[{\rm erg \,s}^{-1}]=46.13$--46.81, which is consistent with the requirement of probing sources capable of supporting radiation-line-driven winds \citep{vms+13,ZG14}.
Four of the 50 targets are known BAL quasars (as identified by \citealt{Shen2011}).  
The full VLA sample is presented in Table~\ref{tab:example}.
%[Update based on \citet{Rankine+2020} BIs?]

%GTR: Moved to here so that the Table numbers are correct.  It will likely appear as a stub in the journal, so it won't take up quite as much space as it seems.

%/Users/gtr/Dropbox/Hewett/Dec2017
%awk -F "," 'NR>1 {print $13 " & " $14 " & " $15 " & " $3 " & " $20 " & " $27 " & " $21}' newPG_DR7.csv } | sort
% awk -F "," 'NR>1 {printf "%s & %.6f & %.6f & %.4f & %.2f & %.2f & %d \\\\ \n",$13,$14,$15,$3,$20,$27,$21}' newPG_DR7.csv } | sort

%see these files to make table below
%https://github.com/RichardsGroup/VLA2018b/tree/master/TABLE_PAPER

\begin{deluxetable*}{LDDcCcccc}
\label{table:VLAsample}
\decimals
%\tablenum{1}
\tablecaption{Target Quasar Properties\label{tab:example}}
\tablewidth{0pt}
\tablehead{
\colhead{Name} & \twocolhead{RA} & \twocolhead{Dec} & \colhead{Redshift} & \colhead{$M_i(z=2)$} & \colhead{$\nu \log L_{1350{\rm \AA}}$} & \colhead{BAL?} & 
\colhead{$S_\mathrm{3GHz}$} & \colhead{$\log L_\mathrm{3GHz}$}\\
\colhead{(SDSSJ)} & \twocolhead{(J2000)} & \twocolhead{(J2000)} & & & 
\colhead{($\mathrm{W}$)} & & \colhead{($\mu$Jy)} & \colhead{($\mathrm{W\ Hz^{-1}}$)}
}
\decimalcolnumbers
\startdata
001342.45$-$002412.6 & 3.426880 & $-$0.403514 & 1.6506 & $-$27.52 & 39.08 & 0 & $163.0\pm17.0$ & 24.47 \\
014023.83+141151.7 & 25.099303 & 14.197709 & 1.6504 & $-$26.09 & 39.21 & 1 & $516.0\pm27.0$ & 24.84 \\ 
014658.21$-$091505.2 & 26.742564 & $-$9.251470 & 1.6509 & $-$26.34 & 39.05 & 0 & $<30.0$ & $<23.61$ \\ 
015720.27$-$093809.1 & 29.334464 & $-$9.635872 & 1.6508 & $-$26.30 & 38.89 & 0 & $<42.9$ & $<23.76$ \\ 
081656.84+492438.1 & 124.236843 & 49.410604 & 1.6508 & $-$26.14 & 39.02 & 0 & $<26.1$ & $<23.55$ \\ 
082334.60+213917.5 & 125.894208 & 21.654881 & 1.6515 & $-$27.32 & 39.51 & 0 & $<31.5$ & $<23.63$ \\ 
082423.61+260656.3 & 126.098409 & 26.115647 & 1.6501 & $-$26.16 & 39.17 & 0 & $<25.8$ & $<23.54$ \\ 
082928.79+401608.4 & 127.369984 & 40.269003 & 1.6507 & $-$26.20 & 38.94 & 0 & $<29.7$ & $<23.60$ \\ 
090502.96+154553.0 & 136.262365 & 15.764731 & 1.6510 & $-$26.17 & 39.03 & 0 & $<60.0$ & $<23.91$ \\ 
092933.99+300240.8 & 142.391663 & 30.044674 & 1.6517 & $-$26.84 & 39.15 & 0 & $83.2\pm11.0$ & 24.05 \\ 
092959.42+270541.6 & 142.497601 & 27.094903 & 1.6515 & $-$27.03 & 39.36 & 0 & $111.4\pm7.5$ & 24.18 \\ 
095648.48+534713.5 & 149.202005 & 53.787094 & 1.6514 & $-$26.58 & 38.82 & 0 & $<36.0$ & $<23.69$ \\ 
100723.31+162842.6 & 151.847151 & 16.478509 & 1.6510 & $-$26.86 & 39.24 & 0 & $<40.5$ & $<23.74$ \\ 
101946.98+494848.6 & 154.945784 & 49.813513 & 1.6513 & $-$26.63 & 39.22 & 0 & $2570.0\pm150.0$ & 25.54 \\ 
104031.38+362611.6 & 160.130766 & 36.436579 & 1.6504 & $-$26.60 & 39.16 & 0 & $100.0\pm12.0$ & 24.13 \\ 
104852.52+003230.0 & 162.218859 & 0.541673 & 1.6506 & $-$27.10 & 39.22 & 0 & $<36.3$ & $<23.69$ \\ 
105814.69+015230.6 & 164.561246 & 1.875186 & 1.6519 & $-$26.45 & 39.08 & 0 & $<660.0$ & $<24.95$ \\ 
110041.31+315018.1 & 165.172157 & 31.838363 & 1.6505 & $-$27.17 & 39.27 & 0 & $294.0\pm17.0$ & 24.60 \\ 
110729.57+270800.7 & 166.873237 & 27.133554 & 1.6501 & $-$26.46 & 39.08 & 1 & $<27.9$ & $<23.57$ \\ 
113047.11+190325.6 & 172.696313 & 19.057130 & 1.6509 & $-$26.40 & 39.19 & 0 & $<25.5$ & $<23.54$ \\ 
113244.74+172629.4 & 173.186458 & 17.441512 & 1.6504 & $-$26.74 & 39.22 & 0 & $63.8\pm6.7$ & 23.93 \\ 
113532.86+001411.4 & 173.886946 & 0.236518 & 1.6517 & $-$26.60 & 39.17 & 0 & $<27.0$ & $<23.56$ \\ 
113548.66$-$022617.8 & 173.952777 & $-$2.438297 & 1.6502 & $-$26.66 & 39.19 & 0 & $<38.4$ & $<23.71$ \\ 
114408.55+020221.2 & 176.035649 & 2.039225 & 1.6506 & $-$26.15 & 39.00 & 0 & $93.0\pm10.0$ & 24.10 \\ 
120015.35+000553.1 & 180.063964 & 0.098102 & 1.6506 & $-$26.40 & 38.88 & 1 & $29.6\pm6.4$ & 23.60 \\ 
120845.22+653344.6 & 182.188432 & 65.562414 & 1.6512 & $-$26.82 & 39.26 & 0 & $<34.5$ & $<23.67$ \\ 
121753.12+294304.6 & 184.471363 & 29.717947 & 1.6507 & $-$26.98 & 39.21 & 0 & $128.1\pm7.4$ & 24.24 \\ 
122225.34+414117.4 & 185.605621 & 41.688191 & 1.6511 & $-$26.67 & 39.20 & 0 & $63.0\pm8.4$ & 23.93 \\ 
124118.12+624606.1 & 190.325501 & 62.768374 & 1.6510 & $-$27.65 & 39.46 & 0 & $316.0\pm18.0$ & 24.63 \\ 
130243.08+130039.3 & 195.679505 & 13.010924 & 1.6505 & $-$27.82 & 39.43 & 1 & $91.5\pm6.1$ & 24.09 \\ 
130545.73+462417.6 & 196.440578 & 46.404898 & 1.6506 & $-$26.71 & 39.12 & 0 & $<25.5$ & $<23.54$ \\ 
132249.49+324826.8 & 200.706248 & 32.807450 & 1.6513 & $-$26.81 & 39.17 & 0 & $<48.3$ & $<23.81$ \\ 
133802.31+084638.4 & 204.509649 & 8.777358 & 1.6502 & $-$26.42 & 39.15 & 0 & $<47.7$ & $<23.81$ \\ 
135348.69+071850.7 & 208.452913 & 7.314084 & 1.6502 & $-$26.51 & 39.19 & 0 & $203.0\pm15.0$ & 24.44 \\ 
140612.82+292303.0 & 211.553449 & 29.384175 & 1.6513 & $-$26.41 & 39.05 & 0 & $<38.4$ & $<23.71$ \\ 
144326.01+305657.6 & 220.858396 & 30.949340 & 1.6513 & $-$26.71 & 39.19 & 0 & $69.5\pm4.4$ & 23.97 \\ 
144510.01+524538.7 & 221.291710 & 52.760759 & 1.6515 & $-$27.23 & 39.38 & 0 & $55.4\pm3.9$ & 23.87 \\ 
144925.90+311041.4 & 222.357935 & 31.178179 & 1.6509 & $-$26.46 & 38.79 & 0 & $55.1\pm8.3$ & 23.87 \\ 
150951.87+075658.8 & 227.466148 & 7.949674 & 1.6517 & $-$26.86 & 39.02 & 0 & $<60.9$ & $<23.91$ \\ 
151450.99+544609.1 & 228.712463 & 54.769198 & 1.6510 & $-$26.49 & 39.22 & 0 & $36.2\pm5.4$ & 23.69 \\ 
153512.10+540215.2 & 233.800417 & 54.037567 & 1.6501 & $-$26.64 & 38.77 & 0 & $50.8\pm2.5$ & 23.83 \\ 
153745.03+481502.9 & 234.437634 & 48.250815 & 1.6517 & $-$26.75 & 39.07 & 0 & $<22.5$ & $<23.48$ \\ 
154631.77+191407.8 & 236.632400 & 19.235513 & 1.6512 & $-$26.58 & 38.98 & 0 & $<28.8$ & $<23.59$ \\ 
160025.19+302751.2 & 240.104995 & 30.464225 & 1.6511 & $-$26.14 & 38.94 & 0 & $<26.1$ & $<23.55$ \\ 
163021.64+411147.0 & 247.590194 & 41.196401 & 1.6511 & $-$27.12 & 39.35 & 0 & $41.4\pm5.2$ & 23.75 \\ 
172057.26+284745.4 & 260.238597 & 28.795948 & 1.6507 & $-$26.28 & 39.06 & 0 & $<19.5$ & $<23.42$ \\ 
213109.57+104714.2 & 322.789885 & 10.787284 & 1.6517 & $-$26.25 & 38.96 & 0 & $<44.1$ & $<23.77$ \\ 
215800.38+002724.2 & 329.501589 & 0.456736 & 1.6504 & $-$26.96 & 39.15 & 0 & $<42.0$ & $<23.75$ \\ 
234607.49$-$002908.7 & 356.531222 & $-$0.485763 & 1.6511 & $-$26.26 & 38.65 & 0 & $<22.5$ & $<23.48$ \\ 
235321.03$-$085930.6 & 358.337649 & $-$8.991838 & 1.6509 & $-$26.73 & 39.09 & 0 & $56300.0\pm150.0$ & 26.88 \\ 
\enddata
%\hline
%\end{tabular}
%\end{table}
\end{deluxetable*}

Two of the 50 potential VLA targets were bright enough ($f_{\nu} > 1\,$mJy\,beam$^{-1}$) to be detected in the FIRST survey \citep{Helfand2015}, while the other 48 required new, deeper radio observations, using the VLA.  We observed these targets in the VLA C configuration and in the S band ($\sim$2--4 GHz) in order to optimize the angular resolution of our images, as well as to leverage archival data from past and ongoing/future radio sky surveys (e.g., FIRST and the VLA Sky Survey [VLASS; \citealt{VLASS}]).  Since our investigation entails testing the possible origins of radio emission, it was necessary to observe down to limits consistent with processes driving the radio emission in radio-quiet quasars.  \citet{mls+16} evaluate the typical break luminosity between star-forming galaxies and AGN in the 1.4-GHz radio luminosity function (RLF) for $z>1$ to be $\log(L_{\rm 1.4GHz}(z)[\mathrm{W\ Hz^{-1}sr^{-1}}]) \approx 21.7+z$.  For the mean redshift of our sample, $z\simeq1.65$, (and changing to units of $\mathrm{W\ Hz^{-1}}$), that corresponds to 
$\log{L_{\rm 1.4GHz}/[\mathrm{W\ Hz^{-1}}]} = 24.45$, which is roughly consistent with other estimates of the break luminosity at this redshift \citep{jr04,Simpson2017}.  Using  a typical radio spectral index for a star-forming galaxy ($\alpha=-0.7$; \citealt{Condon1992}) and aiming to probe at least a factor of five deeper than the observed break in the RLF with a 3-$\sigma$ detection, the desired sensitivity limit was approximately 7.5$\mu$Jy\footnote{In practice 15 of the 28 nondetections had rms sensitivity $>10\mu$Jy, owing to issues such as crowded fields and the Clarke Belt.}.  With 25\% loss of bandwidth due to radio frequency interference (RFI) in the S-band, 28 minutes of on-source observing time per target were required to meet the depth criterion.  Such a depth provided a good balance between the total observing time for a large sample and the ability to attribute nondetections to processes different from those that drive radio emission in radio-loud quasars.  

Similar depth ``pencil-beam" surveys are insufficient to conduct this investigation.  For example, of the 449 sources detected to 7.5$\mu$Jy at 1.4\,GHz by \citet{smg04}, only 1 has a known redshift in the range of $1.5<z<1.8$.  Similarly, of the more than 10,000 sources detected in 2.6 deg$^{2}$ of the COSMOS field at 3\,GHz, only two have $1.6<z<1.7$ \citep{sdz+17}.  Thus, pointed observations such as those presented here have an effective area more than an order of magnitude larger than even moderately wide surveys to deep radio limits.

%[Trevor: what fraction of observations achieved this?] 
%tvm36: This was actually quite rare. Only 3 had sensitivity <=7.5uJy; 21 had <=10uJy; 37 had <= 15uJy
%tvm36: If you want to put it in the context of why we didn't detect as many as we may have thought could also say something like: 15/28 of our non-detections had sensitivity >10uJy owing to crowded fields, Clarke belt, etc...

\subsection{VLA Data Reduction}
We processed the VLA data using the Common Astronomy Software Application \citep[CASA, version 5.4.2;][]{casa}.  Raw data from the VLA were first calibrated by the standard VLA calibration pipeline, and the calibration data later fine tuned by us.  We first reviewed the calibrations of our data from the initial pipeline run to check if any further flagging of our data must be done manually (e.g.,\ for RFI excision or poorly behaved antennas).  If significant flagging of one or more of the calibrators was necessary (as opposed to just flagging the target), we re-ran the pipeline on the flagged dataset.  Once we performed/reviewed preliminary calibrations on our target, we were able to begin the imaging process with the object's fully calibrated measurement set.

To create all images, we utilized the {\tt tclean} task in CASA with natural weighting (which is the weighting scheme that provides the best sensitivity to point sources).  For images that contain other very bright contaminating sources within the field of view, or exhibit clear tropospheric phase errors, we applied up to three rounds of phase-only self-calibration.

While we designed our observations to reach a target sensitivity of 7.5$\mu\mathrm{Jy}$, the actual noise level of an interferometric image depends on weather during the observation, the amount of flagged data, quality of calibration, and the proximity of bright sources within the field-of-view.  For 21 of our targets (13/21 are non-detections), we successfully reached noise levels of $\approx$7.5--10$\mu$Jy; a further 16 targets yielded sensitivity levels of $<15\mu$Jy (with 10/16 undetected).  Of the 11 remaining VLA targets, 10 were constrained to noise levels $<20\mu$Jy, owing mostly to augmented interference in regions coinciding with the Clarke belt.  Four of these 10 are non-detections; for these non-detections a 20$\mu$Jy rms gives a 3-$\sigma_\mathrm{rms}$ upper limit of $\log(L_{\rm 3GHz}[\mathrm{W\,Hz^{-1}}]) = 23.91$.  Finally, the image of SDSS~1058$+$0152 is severely dynamic-range limited, owing to a powerful 120mJy source at 105829.56$+$013401.02 (offset $\simeq18'$ from the central target) and making it difficult to obtain useful radio data for this target.  We nevertheless retain its $\approx220\mu$Jy rms, which corresponds to a 3-$\sigma_\mathrm{rms}$ upper limit on its radio luminosity at $\log(L_{\rm 3GHz}[\mathrm{W\,Hz^{-1}}]) < 24.94$.

\subsection{VLA Detections and Limits}

\begin{figure}[th!]
    %\includegraphics[width=4in,trim= 0cm 0cm 0cm 0cm]{Figures/SDSSRM_Mag_z_Distribution.pdf}
    %\epsscale{1.0}
    %\plotone{Figures/imgGridNew}
    \plotone{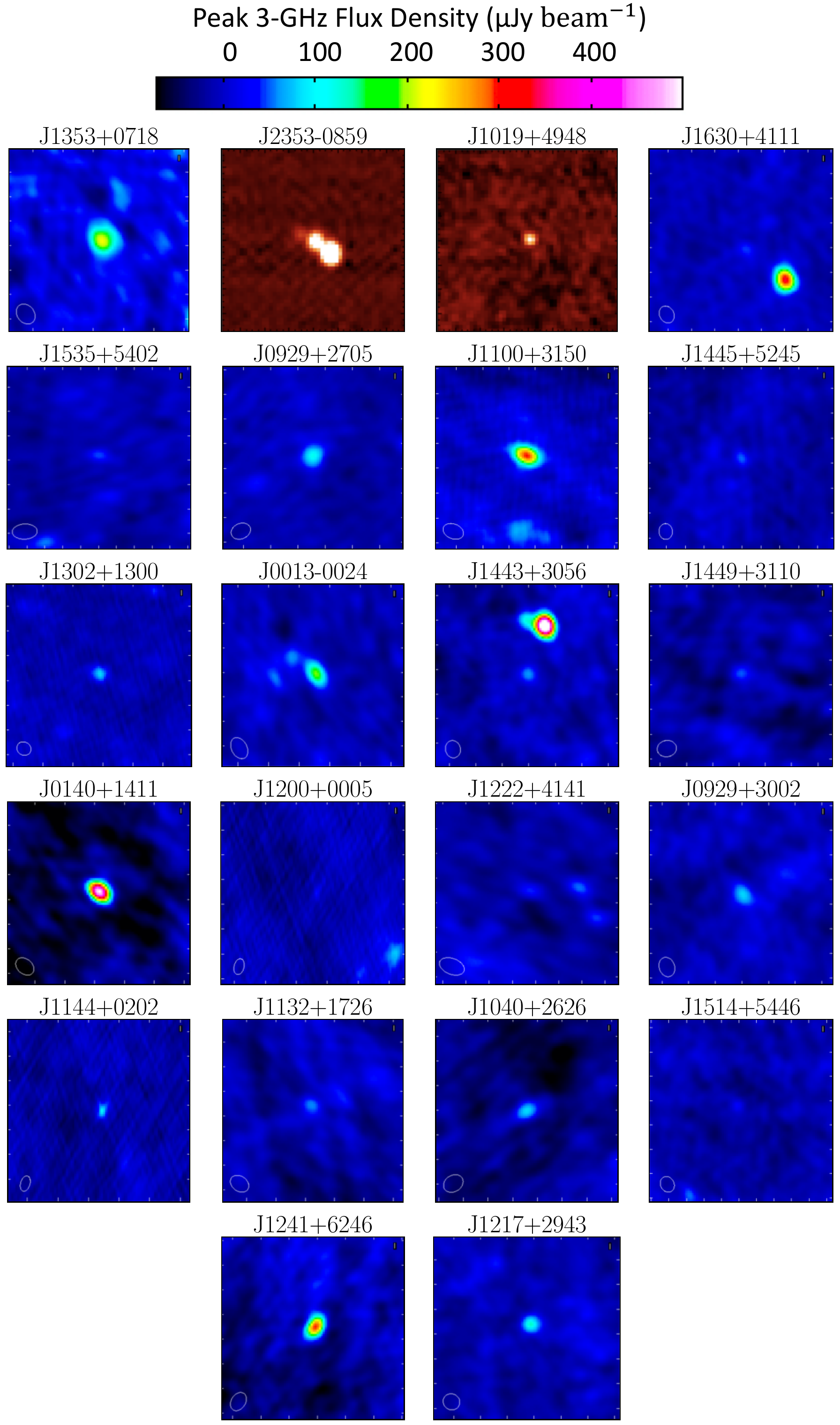}
    \caption{Images of radio detections in our sample ($\simeq90''\times90'')$.  All of the sources detected by our VLA observations (rainbow colormap) at 3\,GHz are on the same linear flux density scale up to $\approx500\mu$Jy, indicated by the colorbar at the top of the figure.  The 1.4-GHz images of the two FIRST sources (red-to-white colormap) are scaled linearly as well, from zero up to the peak flux density of each source ($\approx$50mJy\,beam$^{-1}$ for J2353$-$0859; $\approx$2.5mJy\,beam$^{-1}$ for J1019$+$4948).
    \label{fig:fig1new}}
\end{figure}

\begin{figure*}[th!]
    \epsscale{1.2}
    %\plotone{Figures/nvssForcedPhotometry_GTRoverplotted_wlines}
    \plotone{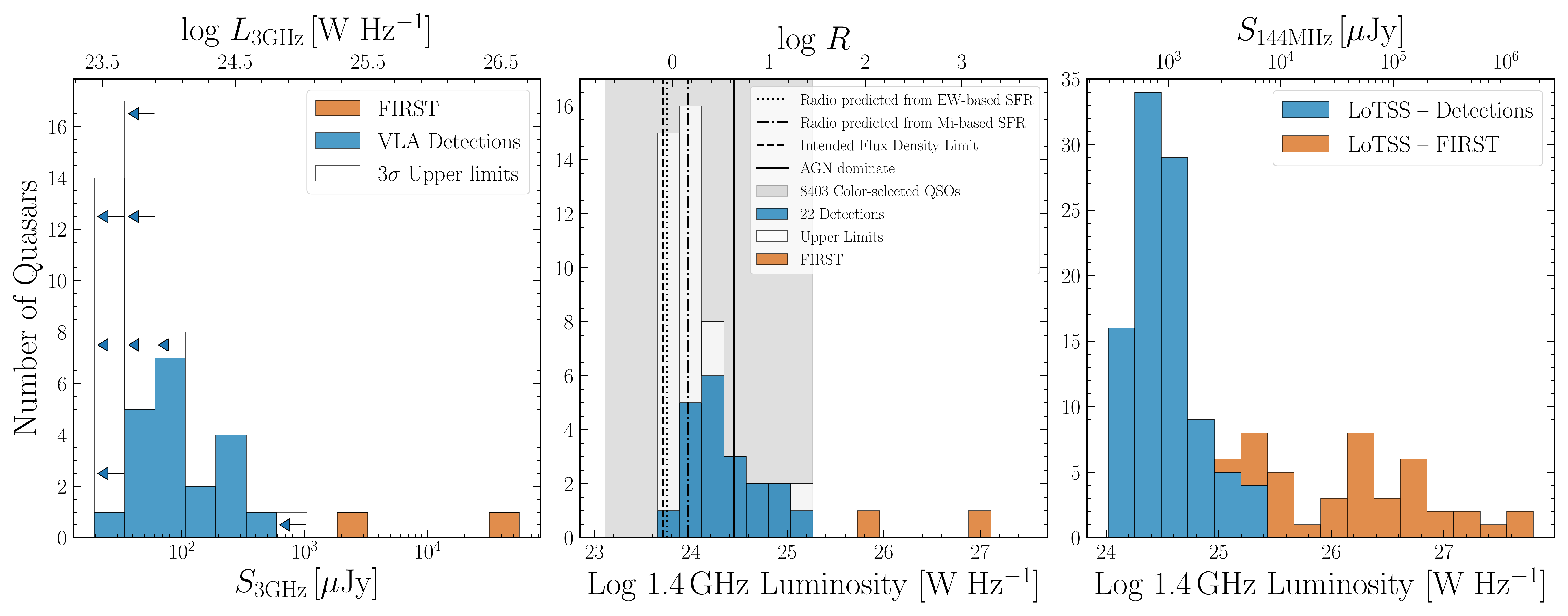}
    \caption{{\em Left:} Distribution of measured total 3-GHz flux densities (blue) and upper limits (open histogram with arrows) in our VLA sample and corresponding luminosity distribution.  Sources with pre-existing FIRST detections are shown in orange.  The top axis converts the measured flux densities to luminosity assuming a source redshift of $z=1.65$.  {\em Middle:} Histogram of radio luminosity for our pointed observations converted to 1.4 GHz assuming $\alpha=-0.7$. The grey shaded region marks $\pm2$-$\sigma$ bounds of the overall distribution of forced photometry on our parent sample (\S\ref{sec:forced}). Vertical lines indicate the expected median radio luminosity from star formation (dash-dot and dotted; see \S\ref{sec:sfr}), the intended radio luminosity depth of our pointed observations (dashed), and the break luminosity at $z\simeq1.65$ (solid; above which AGN contributions to the radio are thought to dominate over SF). The top axis converts the displayed radio luminosity to $\log{R}$ assuming a constant optical luminosity $\log{L_{2500\AA}\mathrm{[W\,Hz^{-1}]}}=23.81$ (the median optical luminosity of the parent sample).  All quasars with $\log{L_{1.4GHz}\mathrm{[W\,Hz^{-1}]}}>25$ have $\log{R}>1$ (the traditional divide between RL and RQ).  {\em Right:}  Stacked histogram of the 123/388 LoTSS-detected sources selected from our parent sample (see \S\ref{sec:LOFARdata}).  Sources detected only in LoTSS are colored blue while those with known FIRST detections are orange. In computing luminosities, measured 144-MHz flux densities are converted to 1.4 GHz assuming a constant radio spectral index $\alpha=-0.7$; for FIRST-detected quasars, we use their measured 1.4-GHz FIRST flux densities.
    \label{fig:fig2new}}
\end{figure*}

Of our 50 targets, 22 are detected at peak flux densities $f_{\nu} > 30\mu$Jy\,beam$^{-1}$, while 3-$\sigma_\mathrm{rms}$ upper limits are obtained for the peak flux densities of the 28 non-detections.  All radio data taken from our sample are reported in Table~\ref{table:VLAsample}.
Images of the 22 detections (including two from FIRST) are shown in Figure~\ref{fig:fig1new}, 
while a histogram of total radio flux density is shown in the left panel of Figure~\ref{fig:fig2new}.  Measured 3-GHz radio luminosities are converted to 1.4 GHz assuming $\alpha=-0.7$ and also to $\log R$ in the middle panel assuming the median optical luminosity of the sample $\log{L_{2500\AA}\mathrm{[W\,Hz^{-1}]}}=23.81$.

%Detection grid made here: https://github.com/RichardsGroup/VLA2018b/blob/master/DetectionGrid.ipynb
%I manually added the colorbar by changing to following "Data Display Options" in CASA: data range=[-7.3e-5, 0.0005], power cycles=-0.3

%See cell 35 of this notebook for how to overplot our 50 sources onto the forced photometry results: https://github.com/RichardsGroup/VLA2018b/blob/master/MakeFigures_forPaper.ipynb

Overall, 12 objects (24\%) are detected at high significance ($\mathrm{S/N}>6$), while the other 10 detections (20\%) are more marginal at $\mathrm{S/N}=3$--6.  
SDSS~2353$-$0859 (a FIRST source) is the only object to exhibit clear signs of extended morphology (what appears to be a core with two lobes, one brighter than the core, one fainter) with just one other object that may have weak extended emission (SDSS~0013$-$0024).  It is possible some sources with wide separation lobes, that are not easily associated with the core, exist.  However, most of the potential candidate lobes either have optical counterparts or would be at separations of $>500$\,kpc.   

We also performed a median stacking analysis on the 28 undetected sources, finding a median peak flux density of 22$\mu$Jy beam$^{-1}$ with rms noise in the stacked image of 2.9$\mu$Jy beam$^{-1}$.  Thus, while the majority of our sources are formally undetected, we would expect them to be detected at $3\sigma$ on average had we been able to achieve our goal of 7.5\,$\mu$Jy rms uniformly across the full sample.

%[GTR: Talk about stacking analysis.  Note that \citet{Condon+2013} argue that the flat radio luminosity function means that stacking results will be dominated by the sources just below the survey limit.  Nadia encouraged us to include this, despite the concerns in Condon+13.]

%tvm36: Final results (in µJy): mean = 8.6 ± 5.9µJy; median = 7.8µJy; peak pixel (which aligns with the final stacked 'source') ~22µJy

%tvm36: With lobes: 1617±20uJy --> L3GHz~25.4W/Hz.
%tvm36: Without lobes: 180±13uJy --> 24.4W/Hz
%tvm36: https://www.evernote.com/l/And9k2kyL29GT72b-0E1iz-fB4l2e-5FLKI

\subsection{Forced Photometry of Parent Sample}
\label{sec:forced}

Our sample of 50 targets was drawn from a parent sample of 8403 known, color-selected, high S/N SDSS quasars as discussed in Section~\ref{sec:targets}.  For the sake of comparison, we performed {\em forced photometry} for the parent sample using the radio images from the NRAO-VLA Sky Survey (NVSS; \citealt{NVSS}), measuring the image peak flux density at the known position of each object.\footnote{https://www.cv.nrao.edu/nvss/NVSSPoint.shtml}  A Gaussian fit to the distribution, sigma-clipping the tail of bright sources, yields a median of 0.09\,mJy beam$^{-1}$ with a standard deviation of $\sigma$ = 0.49\,mJy beam$^{-1}$.  
The error on the median peak is just $\sigma/\sqrt{8403} = 0.005$\,mJy\,beam$^{-1}$, which strongly indicates that the average undetected source has non-zero radio emission.
The range of radio luminosities for this forced photometry is shown by the grey shading in the middle panel of Figure~\ref{fig:fig2new}.
This comparison enables us to see that only the known FIRST sources populate the extreme tail of the distribution that is almost certainly dominated by jetted emission \citep[e.g.,][]{jr04,Simpson2017}.  

\subsection{LOFAR Data}
\label{sec:LOFARdata}

While focused on a very different part of the radio spectrum (144\,MHz) than our VLA S-band observations, the median depth (71\,$\mu$Jy\,beam$^{-1}$ median) of the LoTSS data is similarly sensitive to quasars with radio spectral indices of $\approx -0.7$.  Thus LoTSS provides an additional source of data for our experiment.   We specifically make use of the data from the first data release (DR1; \citealt{Shimwell2019}), covering over 400\,deg$^2$ with optical identifications and morphological classifications provided by \citet{Williams2019}.

\citet{Rankine+2021} demonstrate that radio-loud LOFAR detections (using a defintion of radio loud that is adjusted for the differences in frequency coverage between VLA and LOFAR) are likely drawn from the same parent population as radio-loud sources identified by FIRST.  Thus it is possible to combine the 3GHz and 144MHz data, despite the order of magnitude difference in frequency.   Of the 8403 color-selected sources in our parent sample, there are 388 sources for which LoTSS DR1 should reach comparable depth (for a typical radio spectral index).  Of those 388, 123 sources are detected.  These detections are shown in the right-hand panel of Figure~\ref{fig:fig2new}.  Sources with existing FIRST detections are shown in orange as for the VLA data.

While the LoTSS DR1 sample covers only a small area of sky, the detection fractions for both our VLA observations and in the LoTSS area make it clear that, for the immediate future (until the Square Kilometer Array precursors\footnote{https://www.skatelescope.org/precursors-pathfinders-design-studies/} begin
data releases), pointed observations will continued to be needed for investigations into the origin of radio emission in RQ quasars.  Neither VLASS nor LoTSS reach depths that enable detections of the majority of $z\approx1.5$--2
sources that can best probe a wind origin for radio emission and do so for large areas of sky.

\section{Defining the \civ\ Parameter Space}
\label{sec:c4space}

What we seek to learn herein is to what extent the radio properties of quasars are uniform across a parameter space sensitive to the presence of accretion disk winds, or if the radio properties are systematically changing across that space.  Ideally the parameter space would be defined by physical properties such as black-hole mass, accretion rate, spin, and orientation.  However, each of those properties is difficult to determine robustly for individual objects.  Thus, we adopt the approach of \citet{Rivera+2020} and \citet{Rankine+2021} and explore the radio properties in an empirical parameter space characterized by the \civ\,$\lambda$1550  emission line.

\begin{figure}[h!]
    \epsscale{1.05}
    %\plotone{Figures/radioTargets_1DCIVSpace_wprojection.pdf}
    \plotone{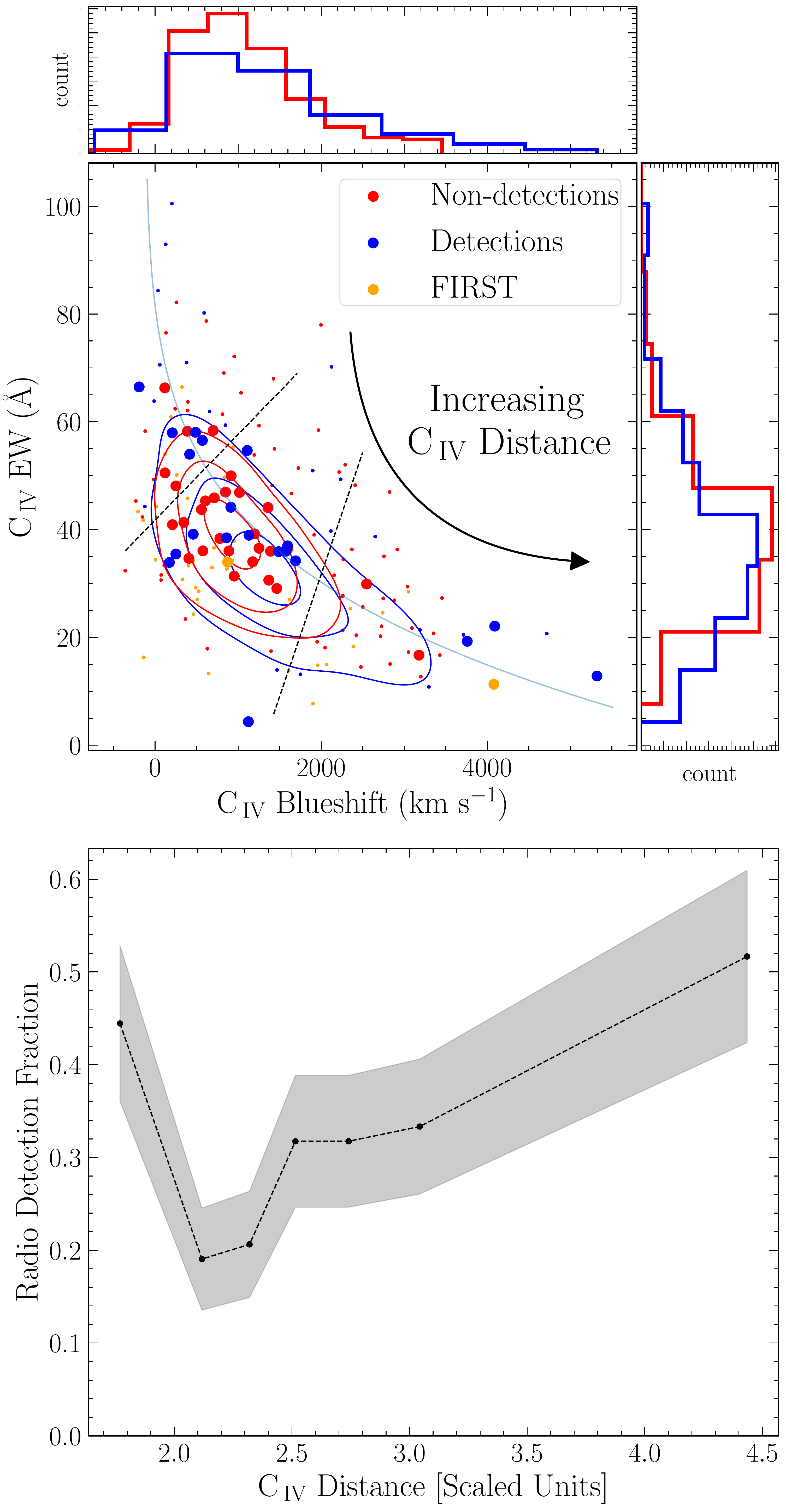}
    \caption{\textit{Top}: Distribution of sources in \civ\ parameter space.  Larger points represent the 50 targets observed with the VLA while smaller points and contours plot the 388 LoTSS objects as discussed in \S\ref{sec:LOFARdata}.  
    The location of each point's orthogonal projection onto the best-fit line (light blue; see \citealt{Rivera+2020}) defines a ``\civ\ distance" metric that we use in comparing each object's location in \civ\ space (see text in \S\ref{sec:c4space}).  Dashed lines (perpendicular to the curve in a scaled data space) show an illustrative division into low, moderate, and high \civ\ distance regimes.  \textit{Bottom}: Radio detection fraction as a function of \civ\ distance.  Fractions are computed in seven bins with equal numbers of quasars, including both the VLA and LoTSS objects; 1$\sigma$ Poisson errors are shown by the shaded region.
    }
    \label{fig:fig3new}
\end{figure}

At low redshift, the principle component analysis (PCA) of \citet{Boroson1992}---primarily involving the H$\beta$, \oiii\, and \ion{Fe}{2} emission lines---concisely describes the diversity of the quasar population in a 2-D space and this type of analysis has been extended by many authors \citep[e.g.,][]{bf99,mds+18}.  At high redshift, the \civ\ emission line by itself defines a 2-D space---equivalent width (EW) and blueshift (a measure of line asymmetry)---that can be used to empirically distinguish quasars with very different physical properties, even without measuring those properties directly \citep{Richards2011,Rivera+2020}.  
Thus we ask how the radio properties are changing across as a function of \civ\ EW and blueshift, guided by expectations as to how the extrema in this parameter space translate to extrema in physical properties.  A key assumption being that the anti-correlatied behavior of the \civ\ EW and blueshift (Figure~\ref{fig:fig3new}) are indicative of the strength of an accretion disk wind.

%See cell 22 of this notebook to plot the CIV plot with the projections; the size scaling for this plot is done in cell 14: https://github.com/RichardsGroup/VLA2018b/blob/master/MakeFigures_forPaper.ipynb
%Find the "CIV distance"-related arrays needed to plot this figure here: https://github.com/RichardsGroup/VLA2018b/tree/master/RM_CIV_Ordering/CIV_PlotArrays
%See this notebook for how we got those arrays: https://github.com/RichardsGroup/VLA2018b/blob/master/RM_CIV_Ordering/RM_CIV_fit_wICA_v2.ipynb

Core to our analysis is spectral reconstruction based on independent component analysis (ICA) as discussed in more detail by \citet{Allen2011}, \citet{Rankine+2020}, and \citet{Rivera+2020}.  ICA serves as a tool to provide a nearly noise-free reconstruction of the spectral features, enabling the most robust measurements of the \civ\ line parameters.  We use the ICA components adopted in the analysis of \citet{Rankine+2020} and compare our results to a similar analysis of a sample of 133 diverse quasars from the SDSS ``Reverberation Mapping" program (SDSS-RM; \citealt{SDSSRM}), as investigated by \citet{Rivera+2020}.

Just as a spectrum can be reconstructed using PCA eigenvectors that are common to a sample with eigenvalues specific to each object \citep{Francis+1992,Yip+04}, so too can a spectrum be reconstructed using ICA ``components" and the component ``weights" that are specific to that object.  The ICA components derived by \citet{Rankine+2020}, analyzed in the same manner as \citet{Rivera+2020}, are employed and the Appendix presents the spectral reconstructions for all 50 quasars in the VLA sample.
We use these reconstructions to extract the EW and blueshift from the spectra (both the VLA and LoTSS quasars) in a manner that results in higher S/N than measurements made directly from the SDSS spectra themselves (as a result of the reconstructions incorporating information from the entire optical/UV spectrum, rather than just the \civ\ emission line itself).  

Figure~\ref{fig:fig3new} shows the distribution of \civ\ EW and blueshift for three sources of data: our color-selected VLA targets (large blue/red points), color-selected LoTSS quasars (blue/red contours and small points) and the SDSS-RM sample from \citet{Rivera+2020}, as represented by the light blue track.
Throughout we will refer to the ``\civ\ distance", which we define using the best-fit curve from the SDSS-RM sample as a ``control", projecting each point in the VLA and LoTSS samples to the nearest orthogonal location on the curve.  Quasars that project onto the start of the curve in the upper-left-hand corner are defined to have zero \civ\ distance, with increasing values of \civ\ distance towards the bottom right.
Thus, high-blueshift quasars have large \civ\ distance, with increasing blueshift being correlated with Eddington ratio \citep[][Fig.~14]{Rankine+2020}.  The top and side panels show the marginal distributions, while the bottom panel combines the EW and blueshift information to reveal that the radio detection fraction is a non-linear function of \civ\ distance. 

It is instructive to look at the data more closely to see if the two subsamples find the same non-linear trend.  The overall radio-detection fraction for our two color-selected samples is 33.1$\pm$2.8\% and for VLA and LoTSS objects separately are 44.0$\pm$9.4\% and 31.7$\pm$2.9\%, respectively.  
At low-blueshift and high-EW (\civ\ distance less than 2.0) we find that the detection fraction is instead 41.7$\pm$7.6\% and is higher for both the VLA and LoTSS quasars separately (50$\pm$22\% and 41$\pm$8\%, respectively).  

While those detection fractions are less than 2$\sigma$ above the average, moving to moderate blueshift and EW (moderate \civ\ distance), we find a significantly lower radio-detection fraction.  Overall it drops to 27.0$\pm$3.0\% and is lower for both the VLA and LoTSS quasars (38$\pm$11\% and 26$\pm$3\%, respectively).

Then moving to high blueshift and low EW (\civ\ distance larger than 3.2), the radio-detection fraction again increases.  The overall fraction is 50.0$\pm$8.5\% and is higher for both the VLA and LoTSS quasars (67$\pm$33\% and 48$\pm$9\%, respectively).  These results are summarized in Table~\ref{tab:det_fractions}.

As the goal of this work is to probe the origin of radio emission in RQ quasars, it is important to also consider these fractions after excluding FIRST sources, which are likely to be dominated by jet emission at the redshifts probed herein.  Doing so leaves the overall trend in radio-detection fraction unchanged: 35.4$\pm$7.4\% at small \civ\ distance, 24.5$\pm$3.0\% at moderate \civ\ distance and 45.0$\pm$8.7\% at large \civ\ distance.  Again the trend is consistent for both the VLA and LoTSS data.  

An obvious concern is that dust-reddening could be playing a role in the radio-detection fraction given known trends between radio emission and both intrinsic quasar colors in the optical and the presence of dust reddening \citep{Richards2003, White2007, Kratzer2015, Klindt2019, Fawcett2020, Rosario2020}.
The color-selected nature of our samples should mitigate against a bias towards more easily finding RL quasars in dust-reddened sources explicitly due to being targeted as radio sources.  However, we find that the trends with \civ\ distance remain even if we exclude the most likely dust-reddened sources ($\Delta (g-i)>0.3$; \citealt{Richards2003}): 37.9$\pm$7.6\% at small \civ\ distance, 25.6$\pm$3.0\% at moderate \civ\ distance and 50.0$\pm$8.6\% at large \civ\ distance.

\citet[][Fig.~3]{Rankine+2021} see a similar distribution in radio-detection fraction with blueshift for LoTSS quasars that are not restricted to being color selected---but without a well-defined rise at low values.  This difference (fall then rise of the \civ\ detection fraction with \civ\ distance in our work versus a consistent rise with \civ\ blueshift in \citealt{Rankine+2021}) may point to the \civ\ distance being a better metric.  For example, in addition to being correlated with blueshift at high blueshift, \lledd\ is also seen to be anti-correlated with \civ\ EW \citep{BL04,SL15}.  Indeed, there is other evidence for the type of nonlinearity that the \civ\ distance captures.  For example, \citet{Rankine+2020} find a gradient of \ion{He}{2} EW along what we define as the \civ\ distance, which distinguishes moderate and high EW sources at similar blueshift.  We further note that \citet[][Fig.~13]{White2007} found a similar non-linear distribution in radio loudness as a function of optical color.  These trends will be discussed in more detail in Section~\ref{sec:analysis}.

\begin{deluxetable}{c|c|c|c}[h!]
    \caption{Detection Fractions as a Function of \civ\ Distance}
    \label{tab:det_fractions}
    \tablehead{
    \colhead{\textbf{\civ\ Distance}} & \colhead{\textbf{VLA}} & \colhead{\textbf{LoTSS}} & \colhead{\textbf{Combined}}
    }
      \startdata
      Low & $50\pm22$\% & $41\pm8$\% & $41.7\pm7.6$\% \\
      Moderate & $38\pm11$\% & $26\pm3$\% & $27.0\pm3.0$\% \\
      High & $67\pm33$\% & $48\pm9$\% & $50.0\pm8.5$\% \\
      \enddata
    \tablecomments{Radio detection fractions for VLA (50 quasars) and LoTSS (388) samples.  ``Low" and ``High" regions of \civ\ distance are illustrated by the dashed lines in Figure~\ref{fig:fig3new}, marking points with \civ\ Distance $<2.0$ and $>3.2$, respectively.}
\end{deluxetable}

%\section{Predictions}
\section{Possible Origins of Radio Emission}
\label{sec:predict}

An experiment that fully explores the diversity of quasars and thus the occurence of the four potential sources of radio emission from quasars \citep[e.g.,][]{Panessa+19} might be expected to reveal one (or more) of the following trends with respect to \civ\ distance.

If the radio emission in RQ quasars is dominated by {\em winds}, then we might expect the radio-detection fraction to increase with \civ\ distance---due to objects with stronger winds (as indicated by larger \civ\ blueshifts) being more likely to have shock-related radio emission \citep[e.g.,][]{LBB19}.  The correlation need not be one-to-one, as said radio emission would require the presence of dense material for the wind to run into; therefore not all sources with winds would be expected to exhibit radio emission.
    
If RQ quasars are instead dominated in the radio by {\em coronal emission}, the radio detection fraction would decrease \citep{RL16,LBB19,GP19} with \civ\ distance.  X-rays provide the indicator of the expected direction of the trend assuming that changes in the X-ray are correlated with the overall size of the corona, and that the coronal X-ray and radio emission are correlated.  
%\citet{LBB19} 
%\citealt{Steffen2006}) 
%\citep{Rivera+2020} 
X-ray emission gets weaker with increasing \lledd, both as found empirically \citep{Sulentic+07,Kruczek+11,Timlin+20} and as predicted by accretion disk wind models \citep{GP19}.  Thus, we expect radio emission of coronal origin to result in less radio emission with increasing \civ\ distance.  

In terms of a {\em star formation} origin, a na\"ive prediction might be that there would be no trend in \civ\ distance.  
%\citep{Richards2006b,shf+20} 
At the redshifts investigated herein the parent sample of SDSS quasars
are all brighter than the ``characteristic" luminosity, $L_*$, that defines
the break in the quasar luminosity function \citep{Richards2006b,shf+20}.  That
being the case, we might expect the SDSS-DR7 sample to lack the
diversity of the full AGN population (i.e., objects both brighter and fainter than
$L_*$)---in terms of properties such as host-galaxy morphology and environment---that are potentially related to star formation rate (SFR) and SFR-related radio emission (see Section~\ref{sec:sfr}).
Moreover, all of our targets herein are drawn from a very narrow range of both redshift and optical luminosity.  If, however, star formation is influenced by ``feedback" processes \citep[e.g.,][]{sr98}, including accretion disk winds, then a trend with \civ\ distance might be expected---given the known correlation between accretion-disk winds and \civ\ distance \citep{Richards2011,wwz+11}.
Specifically, if winds generate positive feedback---in the sense of increasing star formation---then a higher radio detection fraction would be expected with increasing \civ\ distance.  For negative feedback, the opposite would be expected.  Of course, there is no {\em a priori} reason why SF and accretion processes would have to be contemporaneous and we are implicitly assuming that the gas content and star-formation efficiencies of quasar hosts remained unchanged with \civ\ distance.
    
As for a possible {\em jet} origin, we do not make a prediction so much as point out that strong radio sources are anti-correlated with \civ\ distance\footnote{Apparently as are radio-quiet---but otherwise high radio luminosity---flat-spectrum sources \citep{Timlin+21b}.}, and that, while our VLA data probe $\approx 20 \times$ deeper than FIRST, we might not expect more than a single source with jetted emission: FIRST's depth was chosen in order to probe the break in the radio luminosity function, below which star formation starts to dominate over the jetted population.  Therefore we do not expect to find significant jet emission among sources that were not already detected by FIRST.
%(since the FIRST depth was set to probe the break in the radio luminosity function, where star formation may start to dominate over the jetted population).
However, we
acknowledge that our observations (indeed most radio observations) lack the resolution to identify small-scale jets.  
In Section~\ref{sec:discussion} we dicuss some insights on 
whether the presence of such small-scale jets may increase or decrease with \civ\ distance.

In short, our experiment is about looking for trends in radio emission with \civ\ distance.  Our VLA pilot study, even when combined with LoTSS-DR1, is, no doubt, too small to definitively answer the question of the origin of radio emission in RQ quasars.  However, in the absence of information about small-scale jets, an increase of the radio-detection fraction with \civ\ distance might suggest that radio emission is produced by either winds or star formation with positive feedback, while a decrease of the radio-detection fraction with \civ\ distance would point to either a coronal origin or star formation origin (with the AGN contributing negative feedback).

\section{Star Formation Rates}
\label{sec:sfr}

\begin{figure}[th!]
    \epsscale{1.14}
    \plotone{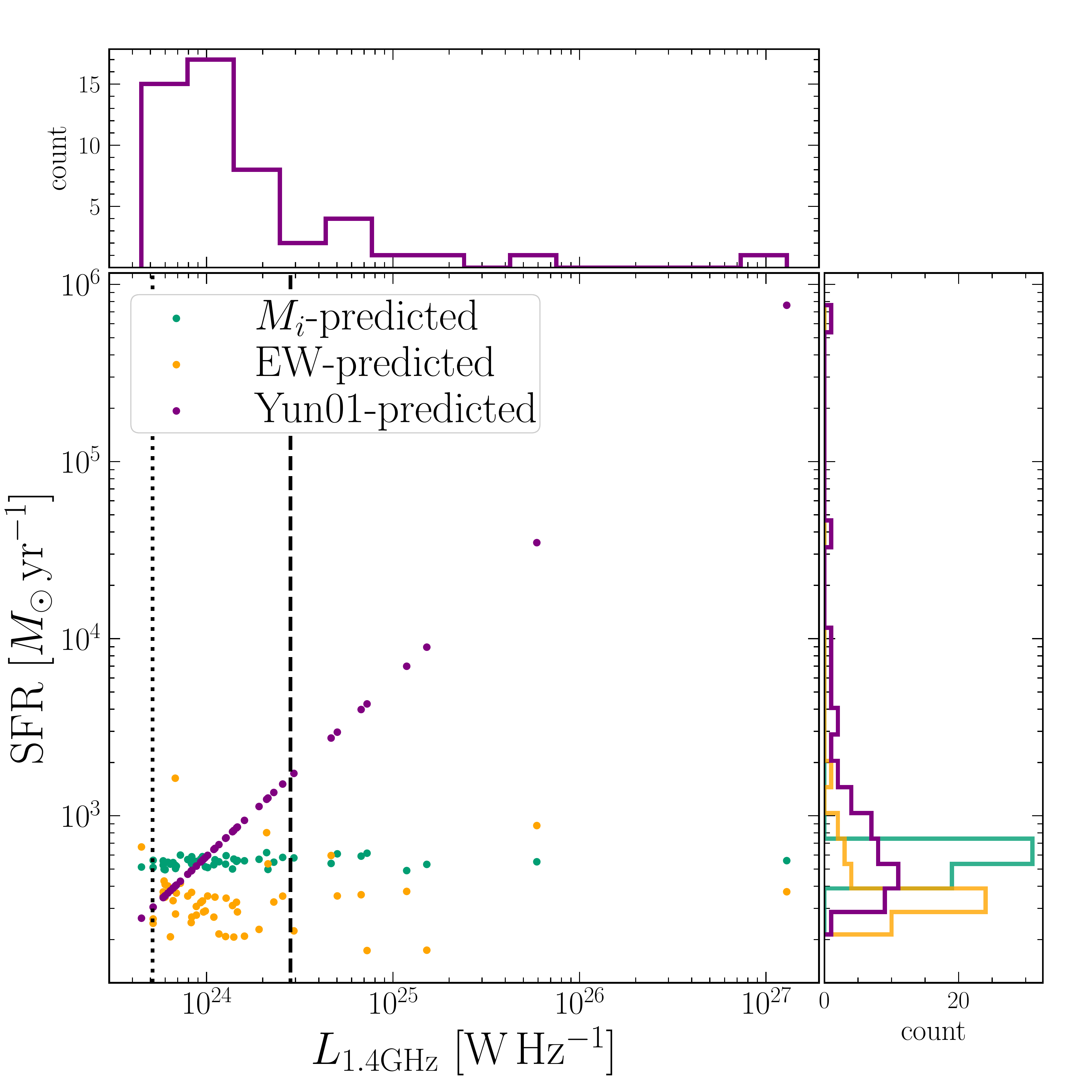}
    %\plottwo{Figures/fig6A}{Figures/fig6B}
    \caption{SFRs of our 50 VLA targets predicted from the observed radio luminosity using \citet{yrc01} and expected based on the optical luminosity [$M_{i}(z=2)$]\ and EW relations from \citet{Harris+16} as a function of their observed radio luminosities (3-GHz flux densities measured from the VLA are converted to 1.4~GHz assuming $\alpha=-0.7$).  Marginal distributions are shown on the top and side axes.  The black vertical lines mark the luminosity above which the AGN is expected to dominate the observed radio emission (dashed) and our intended flux density limit with the VLA (dotted; \S\ref{subsec:VLAExperiment}).
    \label{fig:fig4new}}
\end{figure}

As the predictions in Section~\ref{sec:predict} involve not just expectations for radio emission but also for SFRs, we turn to a discussion of how radio and SFRs are expected and observed to be related in these data.
%GTR: Moving Section 3.7 to here?
We estimate the level of detectable star formation using the far-infrared radio correlation \citep{Helou85,yrc01,cwh+17}, which ties together star formation rates, far-IR emission, and radio emission.  For our VLA observations, we find that a 3-$\sigma$ detection in an image with an rms of 10$\mu$Jy should probe to SFRs of $\sim$400\,$M_{\Sun}\,{\rm yr}^{-1}$.  
In estimating the SFR probed (purple points in Fig.~\ref{fig:fig4new}), we are na\"ively applying the \cite{yrc01} relationship to objects at a very different redshift to that where the correlation was derived.   Furthermore, according to \citet{yrc01}, the relationship between SFR and 1.4-GHz radio applies for objects fainter than $\log(L_{\rm 1.4GHz}[\mathrm{W\,Hz^{-1}}]) \approx 24$.  More accurately, the relation applies to objects below the break of the luminosity function (above which the AGN is thought to be contributing as much or more than star formation to the radio emission).  A break luminosity at $z\simeq1.65$ of %$\log(L_{\rm 1.4GHz}[\mathrm{W\,Hz^{-1}}])
$\log(L_{\rm 1.4GHz}(z)[\mathrm{W\ Hz^{-1}}]) \approx 24.45$ translates to a SFR of $\simeq 1650\,M_{\Sun}\,{\rm yr}^{-1}$.  Thus, radio luminosities (or SFRs) in our sample that are higher should be taken as a strong sign that something other than star formation is contributing to the radio emission, and we indicate this in Figure~\ref{fig:fig2new} as where AGN-related processes might be expected to dominate star formation.\footnote{Noting that an SFR in excess of $\approx 1650\,M_{\Sun}$yr$^{-1}$ is high, but \citet{rwf+18} have argued for the existence of ``extreme" starbursts with SFRs in excess of  $5000\,M_{\Sun}$yr$^{-1}$.}

We can also make specific predictions for radio emission from star formation by taking advantage of far-infrared observations, with the assumption that continuum emission at far-IR wavelengths is dominated by star formation rather than the AGN.  Specifically,
\citet{Harris+16} performed stacking analysis on a sample of 1000 SDSS quasars at $2<z<3$ that are largely undetected by {\em Herschel}.  These data were used to determine the SFR as a function of various properties, under the assumption that the far-infrared emission ($\lambda_{\rm rest} > 80\mu{\rm m}$) is dominated by star formation and not the AGN itself; see also \citet{Rosario+13}.  \citet{Harris+16} present results as a function of the \civ\ EW and optical luminosity, which makes their investigation an excellent source of comparison for our own work.  They also investigated the SFR as a function of \civ\ FWHM and the \civ\ line asymmetry; we have not made use of either of these two results as it is not clear that the FWHM of \civ\ has a clear relationship with black-hole mass \citep{Coatman+17} 
%that is seen in other emission lines (enabling BH mass scaling relations, \citealt{VP06}) 
and the asymmetry parameter is different from the blueshifts investigated herein.  However, \citet{Maddox+17} perform a similar analysis to that of \citet{Harris+16}, measuring the \civ\ blueshifts (using the same definition adopted herein) of a sample of 1185 quasars at $1.6<z<4.8$ with 2-$\sigma$ detections at 
%($\lambda_{\rm obs} > 80\mu{\rm m}$),
$\lambda_{\rm obs} > 250\mu$m (and a matched sample of undetected sources).  Therefore we can compare our work to independent indicators of SFR as a function of both EW \citep{Harris+16} and blueshift \citep{Maddox+17}; i.e., as a function of \civ\ distance.

%tvm36: https://github.com/RichardsGroup/VLA2018b/blob/master/Paper.ipynb

Over a range of more than three orders of magnitude in optical luminosity, \citet{Harris+16} find an increase in SFR from $\approx100$ to $\approx 450\,M_{\Sun}$yr$^{-1}$ with increasing optical luminosity; see also \citet{Stanley+17}.  For the narrow luminosity range of our sample, the expected systematic increase in SFR is from $\approx 490$ to $\approx 620\,M_{\Sun}$yr$^{-1}$.  The SFRs determined by \citet{Harris+16} from binning by \civ\ EW span a larger SFR range ($\approx 50$ to $\approx 500\,M_{\Sun}$yr$^{-1}$) with decreasing EW (from $\approx$ 135\AA\ to $\approx 25$ \AA).  Our sample spans a larger range of EWs than the \citet{Harris+16} sample; the predicted SFRs range from $\approx 190\,M_{\Sun}$yr$^{-1}$ for the largest EW sources ($\approx$ 65\AA) to one object with very low EW (SDSS~J1200+0005) having a predicted value of SFR $\approx 1690\,M_{\Sun}$yr$^{-1}$.  While \citet{Maddox+17} do not perform the same sort of stacking analysis, their comparison of FIR detections to non-detections indicate that quasars with larger \civ\ blueshifts are systematically brighter in the FIR.  Thus objects with large \civ\ distance might be expected to have higher SFRs.
%which is consistent with the anti-correlation between EW and blueshift and the findings of \citet{Harris+16} for SFR as a function of EW.

%Figure~\ref{fig:fig4new} shows the expected 1.4GHz radio emission based on the correlation between SFR and both EW (orange) and optical luminosity (green) found by \citet{Harris+16}, translating the SFR to radio using \citet{yrc01}.
Figure~\ref{fig:fig4new} shows the expected SFR based on the correlation between SFR and both EW (orange) and optical luminosity (green) found by \citet{Harris+16}---against the observed radio luminosity.  
With the caveat that these calculations intrinsically assume that all quasars are similar to the mean quasar, comparing the green and orange points (with SFRs predicted from non-radio data) to the purple points (with SFRs predicted from the radio luminosity using \citealt{yrc01}) reveals that the most luminous radio sources may have contributions to the radio in excess of that expected from star formation.
%translating the SFR to radio using \citet{yrc01}.
However, we can use the results of \citet{Rankine+2021} to understand what might be expected from analysis of a {\em distribution} of quasars instead of considering mean quasar properties via stacking analysis.  Based on detections and non-detections from LoTSS, \citet{Rankine+2021} performed a Monte Carlo simulation that suggests that the radio observations are consistent with a population of SFRs from 10s to 1000s of solar masses per year that could potentially produce radio emission to levels as high as $\log(L_{\rm 144MHz}[\mathrm{W\,Hz^{-1}}]) \approx 26$, which is equivalent to $\log(L_{\rm 1.4GHz}[\mathrm{W\,Hz^{-1}}]) \approx 25.3$ for $\alpha=-0.7$.  It may be that \citet{Harris+16} do not find SFRs in the 1000s as a result of the small area of sky sampled by {\em Herschel} in their study (i.e., a lack of sensitivity to rare quasars with high SFRs), 
%($\sim 1000 M_{\Sun}$) 
and thus that their stacking approach causes us to underestimate the range of SFRs using their empirical correlations.  Therefore, even given the apparent discrepancy with the \citet{yrc01} SFR predictions starting at $\log(L_{\rm 1.4GHz}[\mathrm{W\,Hz^{-1}}]) \approx 24$, it is quite possible that star formation accounts for the radio emission up to $\log(L_{\rm 1.4GHz}[\mathrm{W\,Hz^{-1}}]) \approx 24.5$--25.3. 

\begin{figure*}[th!]
    \epsscale{1.01}
    %\plottwo{Figures/CIVPlot_sizeHarris_SFRinset}{Figures/CIVPlot_sizeRatio_SFRwlinesInset_NDnotscaled}
    \plottwo{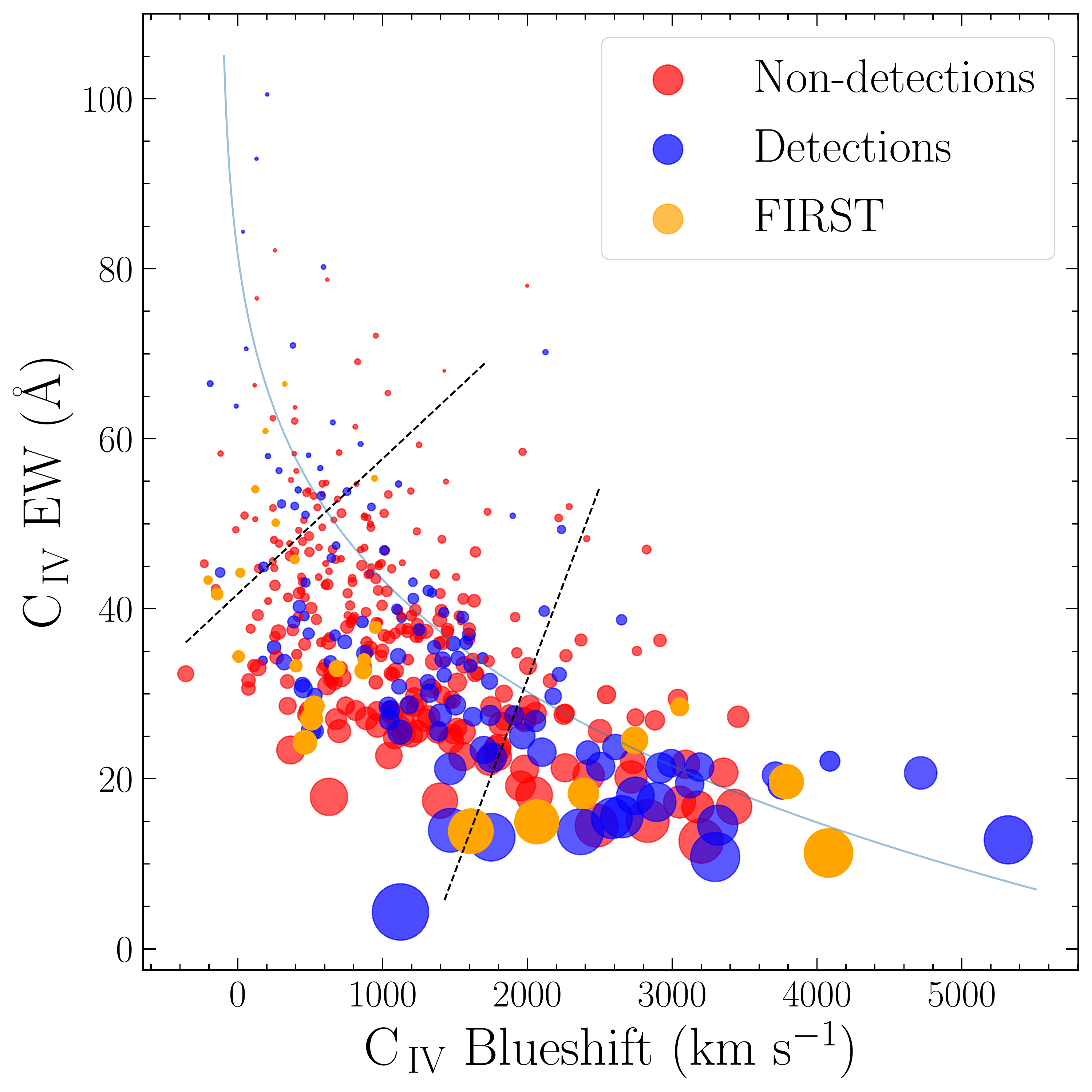}{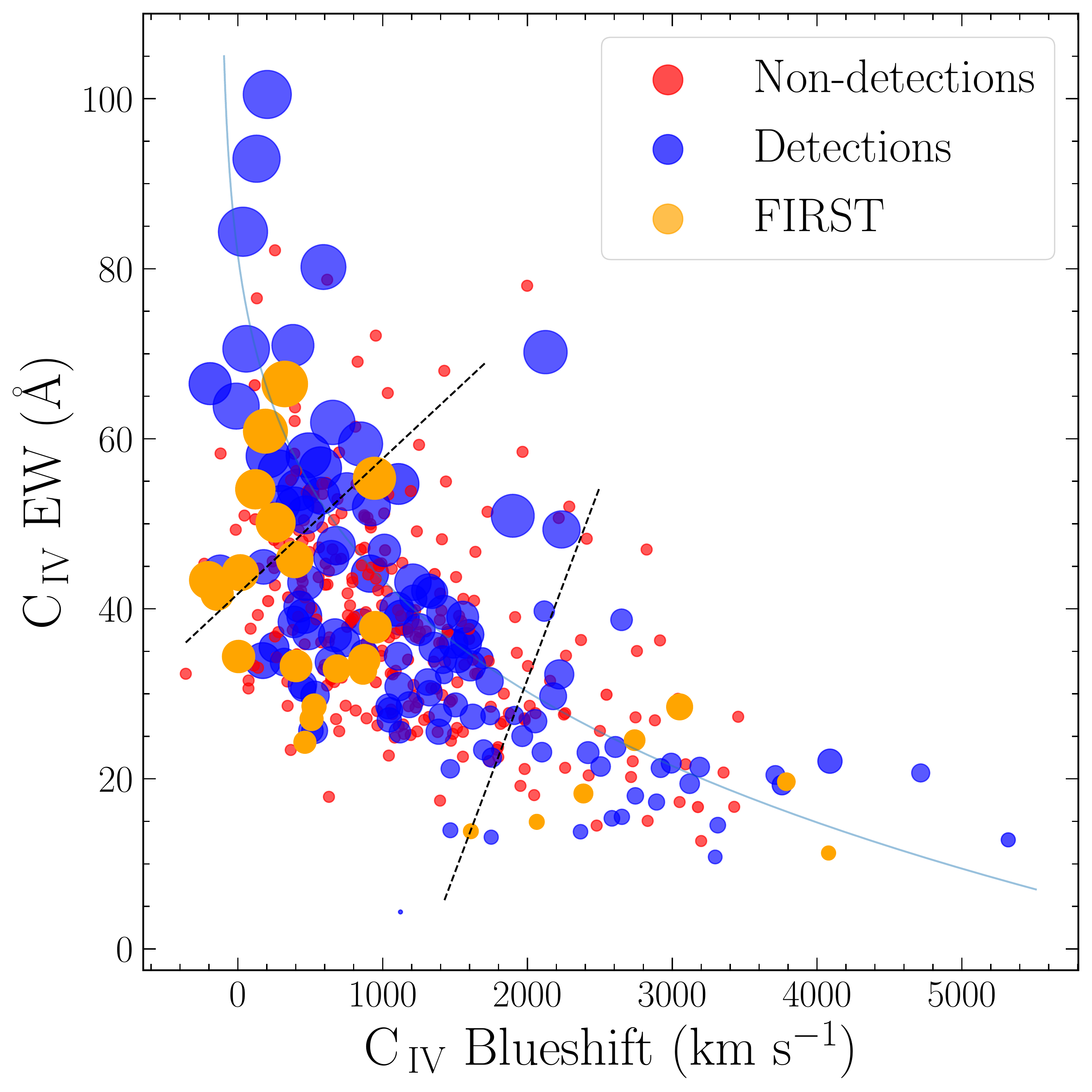}
    \caption{{\em Left:} \civ\ plot with points scaled by SFR predicted from the average of the optical luminosity [$M_{i}(z=2)$]\ and EW relations from \citet{Harris+16}.  
    {\em Right:} \civ\ plot with points scaled by the ratio of observed radio luminosity to that predicted by the \citet{Harris+16} and \citet{yrc01} relationships.  Point sizes larger than that of the non-detections indicate more radio emission than predicted from SF; the {\em detected} quasars are generally brighter in the radio than expected from SF.  %The inset shows the predicted SFRs (based on \citealt{Harris+16} and \citealt{yrc01}) as a function of 1.4 GHz luminosity.  
    The color scheme of both panels follows that of Figure~\ref{fig:fig3new}.  Dashed lines (which are perpendicular to the best-fit curve in scaled space) divide the parameter space into regions of low (upper left), moderate (lower left), and high (lower right) \civ\ distance (see \S\ref{sec:c4space}). 
    \label{fig:fig5new}}
\end{figure*}

Figure~\ref{fig:fig5new} (left) scales the points in the \civ\ space of Figure~\ref{fig:fig3new} by the {\em expected} SFR as predicted by \citet{Harris+16}.  Here we have averaged the SFRs predicted from both the EW and optical luminosity.
It is apparent that the results of \citet{Harris+16} and \citet{Maddox+17} would predict a trend of increasing radio luminosity (from star formation) with \civ\ distance. %(lower EW and higher blueshift).

%To plot just CIV scaled by SFR_obs/SFR_Harris, use 'CIVPlot_sizeRatio_NDnotscaled.pdf'
%To plot the inset of scatters from this plot, use 'CIVPlot_SFRwlinesInset.pdf'
%To plot both together, use 'CIVPlot_sizeRatio_SFRwlinesInset_NDnotscaled.pdf'

%See this notebook for the following plot: https://github.com/RichardsGroup/VLA2018b/blob/master/MakeFigures_forPaper.ipynb
%Left-hand main figure and its inset in cells 25 and 26; size-scaling done in cell 11
%Right-hand main figure and its inset in cells 29 and 30; size-scaling done in cell 11

On the other hand, there is also reason to think that the SFRs estimated by the \citet{Harris+16} analysis are instead over-estimates.  As noted by \citet{Harris+16}, their assumption of an SFR origin for the entirety of the FIR emission ignores the possible kpc-scale dust distribution in AGNs.  Nevertheless, they conclude that AGNs are likely to contribute significantly only to the shortest-wavelength {\em Herschel} bandpass.
A possible reason that this modeling may be incomplete is the assumption of a single SED to describe all AGN, whereas it is well known that AGNs can have significantly different SEDs (at least at 10--20$\mu$m) as a function of optical luminosity \citep{Krawczyk+13}.

\citet{wxz+13} and \citet{Temple+2021} find a significant correlation between \civ\ blueshift and {\em hot} dust, attributing that correlation to a greater ability to view the hot dust component in quasars with strong winds.  If that hot dust component is correlated with a cooler dust component on larger scales, then the assumption inherent to the \citet{Harris+16} argument may not be true.  In such a case, it is possible that the FIR emission is a stronger indicator of AGN-related processes than SFR-related processes; see also \citet{Sym17}.    

While \citet{Sym+16} find no significant difference between low- and high-luminosity AGNs in the FIR, we note that the objects used in that investigation, while luminous for their redshift, are much less luminous than the samples investigated in \cite{Harris+16} and herein.  Thus, the \citet{Sym+16} results do not necessarily contradict the finding of higher FIR emission in more luminous quasars. 

The predicted trend of SFR with \civ\ distance is such that, if the FIR is indeed due to star formation, then quasar winds are either correlated with the same physics that regulates star formation or are providing {\em positive} feedback since those objects with evidence of strong winds are correlated with higher rates of star formation.
If, on the other hand, the results of \citet{Temple+2021} and \citet{Sym17} lead to the conclusion that the FIR emission is AGN dominated, then the predictions for radio emission from star formation derived by \citet{Harris+16} must be considered upper limits (and the prediction of increasing SFR with \civ\ distance is then uncertain).

\section{Discussion}
\label{sec:discussion}

%See notes in "Radio Paper Status, 15 June 2020" in Hewett Evernote folder
Having made predictions for radio emission and SFRs as a function of \civ\ distance, what do we find?  We explore this question in terms of the radio detection fraction, the radio luminosity, and the radio luminosity relative to that expected from SFR---as a function of \civ\ distance (Section~\ref{sec:analysis}).
With our basic results in mind (no simple linear correlation between radio luminosity or detection fraction with \civ\ distance and radio detections in excess of that expected from star formation at low \civ\ distance), we discuss recent work from the literature (Section~\ref{sec:lit}) in an attempt to frame an explanation for our findings (Section~\ref{sec:us}).

%\section{Constraints on the Origin of Radio Emission}
\label{sec:analysis}

\subsection{Results}
\label{sec:analysis}

The non-linear trend in radio detection fraction with \civ\ distance seen in Section~\ref{sec:c4space} does not itself reveal the origin of radio emission in RQ quasars, but it does strongly suggest that there is more than one source of radio emission in RQ quasars, as the observations show a more complex relationship with \civ\ distance than predicted from star formation.  In terms of our predictions from Section~\ref{sec:predict}, the initial decline in radio detection fraction could come from negative feedback on star formation, the corona, or decreasing presence of weak jets.  The subsequent rise in radio-detection fraction could come from positive feedback on star formation or shock-related emission due to winds.      
More work is needed to explore trends in the radio-loud fraction (e.g., as explored by \citealt{Jiang2007}, \citealt{White2007}, \citealt{Kratzer2015} and \citealt{Rankine+2021}) with \civ\ distance to determine if it also follows a non-linear distribution. 

% what are the rms of these two undetected high blueshift sources Perhaps they simply aren't as deep?
%10.5 and 6.5? Is that right? So one is really deep (1720+2847).
%That's correct I remember triple checking the deeper observation. The other high-blueshift source (0823+2139) may actually be a marginal detection
% See slides 15-19 of this slideshow: https://docs.google.com/presentation/d/1KyfniSMB7-_-n00TBLzxs9d2JbaRrcnjbVy6X8_0xuQ/edit#slide=id.g530468978b_0_127

Unlike the trend in radio-detection fraction, we do not observe any clear trends with \civ\ distance in terms of radio luminosity.  However, it is instructive to explore how the radio luminosity relative to that expected from star formation (Fig.~\ref{fig:fig5new}, left) changes as a function of \civ\ distance.
In the right panel of Figure~\ref{fig:fig5new} we instead plot sources according to the ratio of their measured radio emission (for detections) as compared to their predicted radio emission (from the left panel), where points larger than the non-detections (red) are indicative of radio emission in excess of star formation.
While the \citet{Harris+16} predictions only capture the median SFR, there is a systematic trend that may nevertheless be robust.  Specfically, small \civ\ distance quasars have radio emission in excess of that predicted from star formation, whereas quasars at large \civ\ distance (thought to have higher \lledd) are more consistent with SFR predictions of radio emission.  

\begin{figure}[th!]
    \epsscale{1.2}
    %\plotone{Figures/radioVSopt_whistos}
    \plotone{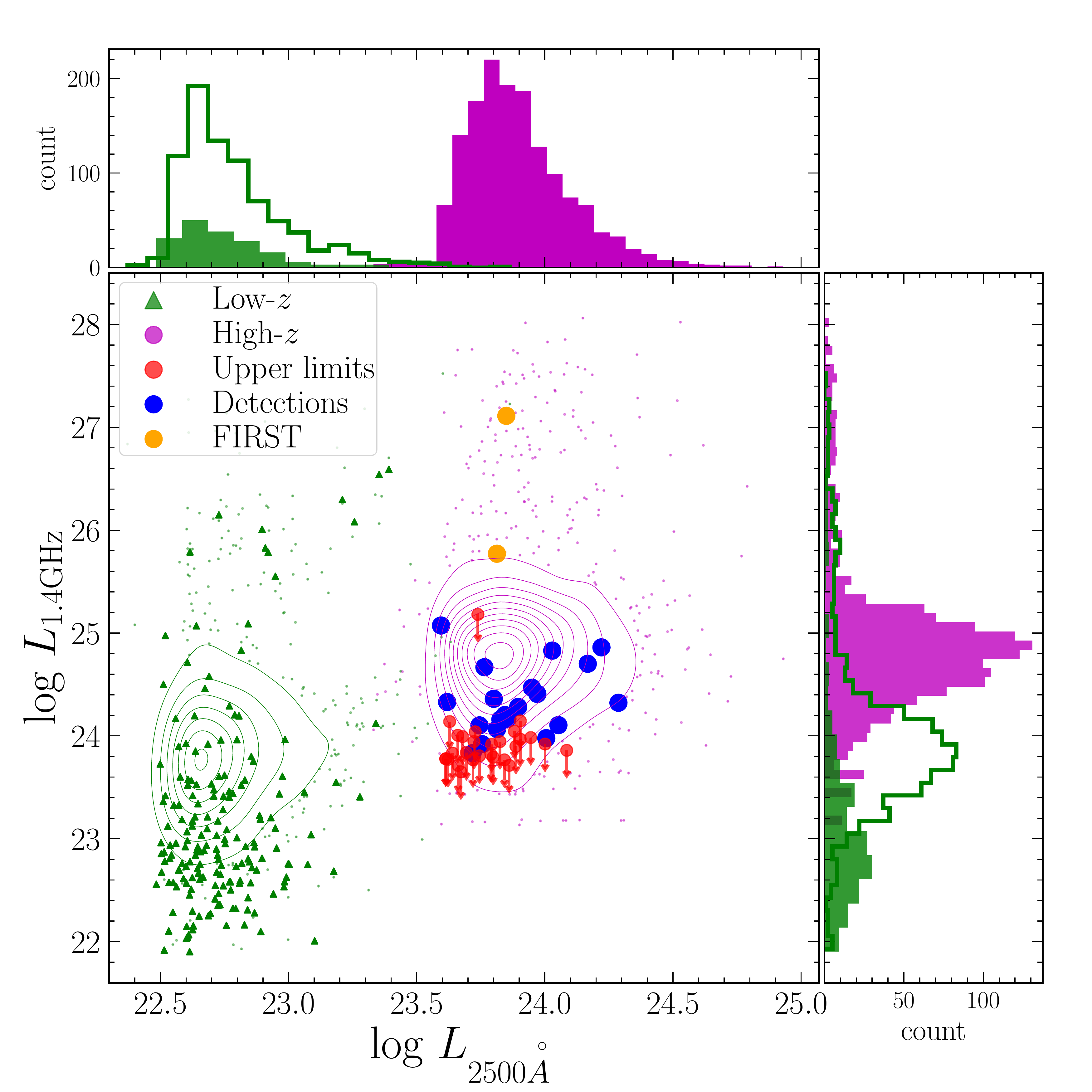}
    \caption{Radio vs optical luminosity (both in W\,Hz$^{-1}$).  Green triangles and filled green histograms represent the full low-redshift ($0.2<z<0.3$) 6-GHz sample of 179 quasars from \citet{Kimball+2011} converted to 1.4\,GHz assuming $\alpha=-0.7$.  \citet{Condon+2013} uses NVSS forced photometry to extend this sample to $z<0.45$; sources with positive flux densities from this sample (965/1313) are shown by the green contours/small data points and open green histograms. Magenta points and contours represent the forced photometry of our parent sample of 8403 quasars (\S\ref{sec:forced}), limited to $z<1.75$ and sources with positive NVSS flux densities, which brings the total source count to 1498.  Our 3-GHz VLA data of 50 quasars converted to 1.4\,GHz assuming $\alpha=-0.7$ are shown in red, blue, and orange; the downward arrows on the non-detections emphasize that their radio luminosities are plotted as 3-$\sigma_\mathrm{rms}$ upper limits.  The marginal distributions 
    %of the \citet{Kimball+11b} and high-z data are shown on the axes.
    are given by the histograms on the top and right axes.
    \label{fig:fig6new}}
\end{figure}

If the origin of radio emission is primarily related to the AGN, then one might expect the radio luminosity to scale simply with the optical/UV luminosity.  We test that prediction by comparing our data to that of \citet{Kimball+2011} and \citet{Condon+2013} in Figure~\ref{fig:fig6new}, finding that our sources are about 1.1 dex more luminous in the optical/UV and 1 dex more luminous in the radio.  Of course, it is possible that the observed higher radio luminosity with higher optical luminosity could also be associated with higher star formation \citep{Stanley+17}.
%the trend seen in Figure~\ref{fig:fig6new} might be taken as evidence for an AGN rather than star formation origin for the radio detections in our sample.  
%The opposite may be true 
The radio non-detections, despite having much higher optical/UV luminosity, have upper limits on their radio luminosities that are not inconsistent with those from the lower-redshift sample from \citet{Kimball+2011} and \citet{Condon+2013}.

%See cells 37-44 one how to create the radio vs opt distribution plot: https://github.com/RichardsGroup/VLA2018b/blob/master/MakeFigures_forPaper.ipynb

\subsection{Literature Review}
\label{sec:lit}
 
%[GTR: 16 October 2020] Changed from present to past tense.  I might prefer present as the results are in the past, but presumably are still true today. 
 
\citet{Kimball+2011} analyzed a sample of 179 SDSS quasars with $0.2<z<0.3$ and more luminous than $M_i=-23$.  They concluded that the radio luminosity function is consistent with two independent origins for radio emission in quasars, which they attributed to starburst-level SF processes at the faint end and AGN processes at the bright end.  The transition region is roughly $\log L_{\rm 6 GHz} = 29.5$--$30.5\, {\rm erg \, s^{-1}}$, which corresponds to $\log L_{\rm 1.4GHz} = 23.34$--$24.34\, {\rm W \, Hz}^{-1}$, assuming linear evolution of the break luminosity with redshift.
%(29.5-7)*(1.4/6)**-0.7 + 1.65-0.25 =  24.34 
Figure~\ref{fig:fig6new} demonstrates that the objects in our sample are much more luminous than the lower-redshift sample of \citet{Kimball+2011} in the optical/UV and also the radio.  

\citet{Condon+2013} extended the work of \citet{Kimball+2011} by building a larger low redshift sample (1313 quasars at $0.2<z<0.45$) and adding a high redshift sample (2471 quasars at $1.8<z<2.5$), which is more similar to the sample investigated herein.  
They found that, analogous to the low-redshift sample, there is sign of multiple radio origins among NVSS sources: radio emission in the RL tail is dominated by AGNs, but the radio luminosity distribution must have a bump at low flux densities, which is consistent with a star formation origin dominating the radio emission for the fainter sources.
They argued that the level of radio emission in their high-$z$ sample suggests SFRs as high as $\dot{M} \sim 500 M_\odot \, {\rm yr}^{-1}$. 

On the other hand, \citet{White+17} analyzed a sample intermediate in redshift between the two \citet{Condon+2013} samples, specifically 70 $z\approx 1$ Herschel AGN.  As with \citet{Harris+16} they used FIR emission as a tracer of star formation---assuming that AGNs contribute little at the longest FIR wavelengths.  Using two SFR correlations they found that 92\% of RQ quasars have radio emission that is accretion dominated and that 80\% of the emission is due to accretion.

\citet{Mullaney+13}, \citet{ZG14}, \citet{Zakamska+16}, and \citet{Hwang+2018} similarly concluded that radio emission in RQQs is dominated by the AGN, but argued that the emission is related to shocks from uncollimated, subrelativistic winds interacting with the inter-stellar medium.  Papers addressing such an origin from the theoretical perspective include \citet{Ciotti+2010}, \citet{Jiang+10}, and \citet{Nims+15}.  \citet{Nims+15} argued that it should be possible to distinguish between jets and winds by the spatial extent of emission and radio spectral index (where they suggest $\alpha\approx-1$ for winds).  However, it is harder to then distinguish between AGN and star formation---without looking to other wavelengths for help.

\citet{ZG14} found that the \oiii\ velocity width of outflows is correlated with radio luminosity in RQ quasars and that the width is additionally correlated with mid-IR luminosity, also suggesting an AGN-intrinsic origin (because the mid-IR is dominated by the AGN). 
%[Interestingly they quote an SFR of 400 as being unrealistically high, whereas Condon is apparently OK with that.]  
\citet{ZG14}, however, instead looked to shocks from quasar-driven winds as suggested by \citet{Stocke1992}, similar to how SNe produce radio emission.
They concluded that it would be hard to distinguish jet emission from wind emission based on morphology of the ionized gas or the radio spectral index, 
%(and that would be weird if all RQ have frustrated jets).
but there must be a threshold for wind driving on the order of bolometric luminosity, $L_{\rm Bol} = 3\times10^{45} {\rm erg \, s}^{-1}$.  The median $\log L_{\rm Bol}$ for our targets is 46.3 in those same units---consistent with the design of the observing program to target quasars likely to show evidence of accretion disk winds.

%Range is 46.13 to 46.81

To further connect the \citet{ZG14} result with our own, we note that \citet{Coatman+19} showed that \oiii\ blueshift and EW are correlated and anti-correlated, respectively, with \civ\ blueshift.  Thus, if winds contribute significantly to radio emission in RQ quasars, we might expect a bias towards more/higher radio emission at high blueshift (large \civ\ distance).  While there are indications of winds in the form of BALQSOs across the full \civ\ parameter space \citep{Rankine+2020}, if radio emission is coming from shocks due to winds, then we might expect quasars with the strongest troughs (largest balniticy indices; \citealt{Weymann1991}) to produce more radio emission.  We note, however, that the winds highlighted by \citet{ZG14} are associated with strong \oiii, whereas the strongest BALQSOs tend to have have weak \oiii\ \citep{Turnshek+97}.  Thus it may be that \oiii\ is no longer useful as a tracer of winds for the very objects where it might be most interesting (i.e., those with the strongest winds; compare Figure~8 in \citealt{Coatman+19} and the right column of Figure~16 in \citealt{Rankine+2020}).
One explanation is that the narrow line region (NLR) may be ``running out of gas" at high luminosity \citep{Netzer90,Hainline+14}.  Moreover, only with IR spectroscopy is \oiii\ available at $z\sim2$ redshifts---spanning the peak of SF and AGN activity.  Thus investigation of winds using \civ\ allows for construction of larger samples.  While the winds traced by \oiii\ and \civ\ presumably are at very different distances from the central engine, the NLR wind traced by \oiii\ in the results of \citet{Coatman+19} suggests that the NLR may be connected to winds driven by the central engine, which ultimately influences galaxy scales.
    
\citet{Zakamska+16} analyzed two $z<1$ samples: type-2 objects from \citet{rzs+08} and type-1 objects from \citet{sor+07}.  While they agree with \citet{Kimball+2011} and \citet{Condon+2013} that their results indicate that a second contribution beyond jets is needed, they concluded that the SF is insufficient to explain radio by an order of magnitude.    As a result, \citet{ZG14} and \citet{Zakamska+16} make the case for radio emission resulting from shocks due to uncollimated, sub-relativistic winds impacting the ISM of the host galaxy. 

\citet{Hwang+2018} found that a special population of quasars (extremely red quasars [ERQs] at $2<z<4$) have moderate-luminosity, compact radio sources with steep spectra.  They argued that the excess emission is more likely to be due to uncollimated winds than star formation or coronal emission.
Such outflows can be from radiation line-driven winds, radiation pressure on dust, or magnetic fields--- all of which can produce synchrotron from shocks \citep{Jiang+10,Nims+15}.
\citet{Hwang+2018} concluded that the radio can be from winds if just 1\% of the bolometric luminosity is converted to kinetic energy.
%Steepness could be from cooling (Jiang+10), or maybe Inverse Compton (Nims+15).
%CSS would be unresolved at this resolution, but would be odd if selection of ERQs picked out CSS sources more easily than in the general population.

On the other hand, \citet{Jarvis+19} argued that evidence for a wind origin of radio emission based on \oiii\ may be biased by the lack of high-resolution radio data as they find evidence for jets in their sub-arcsecond resolution images.
However, the luminosity of their sources is below the threshold for winds as determined by \citet{ZG14} and probed by both our targets and the \citet{Hwang+2018} sources.  Nevertheless, other evidence for weak/frustrated jets exists \citep[e.g.,][]{BB98,uab05,hr+16,Nyland+20}, but it is unclear to what extent such systems operate at the redshifts and luminosities investigated herein \citep{Condon+2013}.

\subsection{Multiple Sources of Radio Emission in RQ Quasars?}
\label{sec:us}

Even at the depth probed by our observations, less than 50\% of quasars are detected in the radio, making it difficult to identify clear trends with \civ\ distance that can be used to test the hypotheses that we have outlined (where the stochasticity of the detections may itself be an important clue to the origin of radio emission in RQ quasars).  However, one of the reasons for the lack of a clear trend may be that RQ quasars have multiple sources of radio emission and it may be that those origins have opposite trends in terms of \civ\ distance.

Taken at face value, the results of \citet{Harris+16} and Figure~\ref{fig:fig5new} suggest that quasars with small to moderate \civ\ distance have a significant source of radio emission in excess of that from star formation, which becomes less significant for low-EW (higher \lledd) sources.  Such a source could be the corona (see Section~\ref{sec:predict}).  
%For example, in the picture of \citet{RL16}, this decrease in radio detection fraction might be due to coronal emission being correlated with X-ray emission (which goes down with increasing \civ\ distance as a result of trends with luminosity and the well-known $L_{\rm uv}-\alpha_{\rm ox}$ relationship) and cooling of the corona by the accretion disk.  
For example, in the picture of \citet{RL16}, radio emission from the X-ray corona could correlate with X-ray luminosity. The X-ray luminosity decreases with increasing \civ\ distance as a result of trends with overall luminosity (and the well-known $L_{\rm uv}-\alpha_{\rm ox}$ relationship) and cooling of the corona by the accretion disk. Such a trend could thus reproduce the decrease in radio detection fraction with \civ\ distance at small to moderate \civ\ distance.

This radio excess and the decreasing radio detection fraction with \civ\ distance are inconsistent with the predictions of \citet{Harris+16} and a star formation origin of this radio emission in RQ quasars.
These empirical results are also unlikely to be due to winds (given that the shape of the optical-to-X-ray SED changes across \civ\ space in a way that one would predict weaker winds at high \civ\ EW; e.g., \citealt[][]{GP19}).

As 
%the formally radio-loud population also goes down with \civ\ distance
the prevalence of radio-loud sources decreases with \civ\ distance \citep{Stocke1992,Richards2011,Kratzer2015,Rankine+2021}, we suggest that the presence of jets decreases with \civ\ distance; see also \citet{Timlin+21b}.  If frustrated/weak/small-scale or newly active jets \citep[e.g.,][]{uab05,Nyland+20} will evolve into RL sources, such jets may follow the same trend as RL sources and could account for the excess of radio emission over that expected from star formation at small \civ\ distances.  That would also be consistent with radio detection fraction going down from the lowest to moderate \civ\ blueshifts.

Going from moderate to high \civ\ distance we instead see a trend of increasing radio detection fraction, consistent with the results of \citet{Rankine+2021}.   Given the anti-correlation with the radio-loud fraction and with X-ray emission, it would seem that the source of this trend is unlikely to be weak jets or coronal emission \citep{LBB19}.  Instead it is more consistent with star formation or shocks from winds.  Considering the well-known Baldwin (\citeyear{Baldwin77}) effect and the anti-correlations between \civ\ EW and luminosity, the simplest conclusion may be that the trend of increasing radio detection fraction with \civ\ distance simply reflects more star formation in higher luminosity sources (e.g., \citealt{Stanley+17}, and indirectly illustrated in the left-hand panel of Figure~\ref{fig:fig5new}).  The right-hand panel of Figure~\ref{fig:fig5new} is indeed consistent with that hypothesis as the point sizes indicate that the expected radio emission is broadly consistent with that expected from star formation at large \civ\ distance.

However, we caution that that is not the only interpretation of the agreement between predicted and observed radio emission at large \civ\ distance.  If the IR emission that is crucial to the construction of the SFR estimates in \citet{Harris+16} is contaminated by emission from an AGN,\footnote{As might be suggested by the correlation seen between blueshift and hot dust by \citealt{wxz+13} and \citealt{Temple+2021}---which cannot be due to direct star-formation processes, but could be due to a wind-modulated connection between the hot dust at small distances from the BH and cold dust at large distances.} then the FIR predictions for radio emission break down.   The apparent agreement between predicted and observed radio emission at large \civ\ distance could instead be indicative of radio emission dominated by shocks from winds.  That is, consistency with predictions from star formation does not necessarily confirm a star formation origin for radio emission in RQ quasars at large \civ\ distance.  Indeed, such a trend in \civ\ space is exactly the behaviour we might have expected from a component of radio emission with a wind-shock origin.
As such, it is not clear that trends in \civ\ alone could be used to distinguish between a star formation origin and a wind origin of radio emission.  More observational and theoretical work is needed to determine if the predictions themselves at large \civ\ distance could be influenced by accretion disk winds. 
%Moreover, our work (including \citealt{Kratzer2015} and that of \citealt{Rankine+2021}) suggests that the ratio of radio {\em detections} to radio {\em loud} sources increases systematically with \civ\ distance (in that quasars are more likely to be detected, but less likely to be formally radio loud).  
However, such an observation is most consistent with either a direct wind origin for radio emission, winds contributing positive feedback to star formation, or the radio and star formation being driven by the same underlying physical parameter/conditions.  A wind-related origin could explain both the increasing radio-detection fraction (but decreasing radio-loud fraction) with \civ\ distance and the stochastic nature of radio detection in an otherwise homogeneous sample.  Thus we conclude that a wind-related shock origin for radio emission in RQ quasars merits further investigation. 

\citet{LBB19} make the case that the origin of radio emission in RL and RQ quasars is different as a result of different trends in those two populations.  Our work is consistent with that interpretation and further suggests that, even within the RQ quasar population, there may be multiple sources of radio emission and that the \civ\ distance is a useful diagnostic tool for exploring that hypothesis.
%\citet{Stocke1992}.  %\citet{Richards2011} 
%\citealt{Rivera+2020}), 
%the conclusion that one draws about the likelihood of radio emission at large \civ\ distance being dominated by winds depends strongly on the expected trend of radio emission from star formation.  Thus, more work is needed to understand the extent to which the ``floor" in radio detection fraction is set by star formation and how that changes with \civ\ distance.  
Thus, we must consider the possibility that radio emission in quasars simply does have multiple origins: jets, star formation, coronae, and winds \citep[e.g.,][]{Kimball}, where it may be the case that those processes are anti-correlated in a way that hinders answering the question of the origin of radio emission in RQ quasars.
%and why the majority of our sources are undetected at the expected level for star formation. Answering these questions will require a larger sample of objects (with quasars that are likely to exhibit accretion disk winds) with observations down to much lower SFRs and with high enough resolution to find small-scale jets.
This picture is consistent with the results of \citet{Timlin+21b}, who find that only RQ quasars that are flat-spectrum and have high radio luminosity exhibit excess X-ray emission from jets.  That conclusion is equivalent to saying that only such RQ quasars have their radio emission dominated by jets.  As they find that these sources have large \ion{He}{2} EWs, they also have small \civ\ distance.

\section{Conclusions}
\label{sec:conclusions}

We targeted 50 color-selected individual SDSS quasars at $z\simeq1.65$ for VLA radio observations probing the origin of radio emission in otherwise radio quiet quasars.  We coupled these data with 388 quasars covered by LoTSS DR1 that were drawn from the same parent sample of color-selected quasars. Such luminous, high-redshift quasars provide an important complement to investigations of radio emission in RQ quasars based on evidence for winds from \oiii\ emission as the \civ\ emission line provides a unique diagnostic of winds that are likely to have an origin in radiation line driving.  To aid in this analysis, we define a \civ\ ``distance" that maps both \civ\ EW and blueshift onto a single non-linear parameter and explore the radio properties of RQ quasars as a function of that distance (large EW having small distance and large blueshift having large distance).

We find that only 22 of our targets are detected at radio luminosities in excess of $\log L_{1.4GHz}/[{\rm W \, Hz}^{-1}]$ = 23.71, which corresponds to a star formation rate of $\approx 300\,M_{\Sun}\,{\rm yr}^{-1}$ using the method of \cite{yrc01}.  Similarly only 123 of the 388 LoTSS souces are detected.  For quasars with small \civ\ distance, the detected quasars have radio emission intermediate between that expected from predictions of median star formation and that typically seen in sources with large-scale jets, possibly suggesting coronal emission or weak/small-scale jets. 
There is evidence for a higher radio detection fraction at either extrema of the \civ\ distance, possibly suggesting multiple origins of radio emission in RQ quasars.  Until new surveys come on line, more pointed observations are needed on more quasars to deeper radio limits (and higher resolution) in order to resolve the question of the origin of radio emission in these sources and determine the level to which winds, coronae or weak jets may compete with star formation processes.  That said, the non-detections in our sample have radio emission that is consistent with that predicted from star formation, but only if SFR predictions based on far-IR emission are not biased by AGN-heated dust on galactic scales, which could imply greater contribution from winds.

Future observational work suggested by this investigation would be to determine the \civ\ distances of weak/young jet sources such as those from \citet{Nyland+20}.  Similarly, additional deep, high-resolution observations are needed to test for compact jets.  It would further be useful to explore not just the radio-detection fraction of the parent sample, but the radio-loud fraction as well, since past work has shown those properties to be anti-correlated \citep{Kratzer2015,Rankine+2021}.  However, the tiny fraction of quasars that are RL mean that this requires survey data.  Finally, following \citet{LBB19}, an analysis of spectral indices (or more generally radio SEDs) of the sample (e.g., using VLASS data) would reveal if there are any trends in terms of optically thick vs.\ optically thin radio emission as a function of \civ\ distance that could be used to understand the origin of the radio emission in RQ quasars.

\acknowledgements
We thank Nadia Zakamska for fruitful conversations and Josh Marvil of NRAO for assistance with correction of VLA gain compression.  We thank the referee for suggestions that significantly improved the paper.  ALR acknowledges funding via the award of a Science and Technology Facilities Council (STFC) Ph.D. studentship. PCH acknowledges funding from STFC via the Institute of Astronomy, Cambridge, Consolidated Grant. The National Radio Astronomy Observatory is a facility of the National Science Foundation operated under cooperative agreement by Associated Universities, Inc.  Funding for the SDSS and SDSS-II was provided by the Alfred P. Sloan Foundation, the Participating Institutions, the National Science Foundation, the U.S. Department of Energy, the National Aeronautics and Space Administration, the Japanese Monbukagakusho, the Max Planck Society, and the Higher Education Funding Council for England. The SDSS Web Site is http://www.sdss.org/.  The SDSS is managed by the Astrophysical Research Consortium for the Participating Institutions.  The Participating Institutions were the American Museum of Natural History, Astrophysical Institute Potsdam, University of Basel, University of Cambridge, Case Western Reserve University, University of Chicago, Drexel University, Fermilab, the Institute for Advanced Study, the Japan Participation Group, Johns Hopkins University, the Joint Institute for Nuclear Astrophysics, the Kavli Institute for Particle Astrophysics and Cosmology, the Korean Scientist Group, the Chinese Academy of Sciences (LAMOST), Los Alamos National Laboratory, the Max-Planck-Institute for Astronomy (MPIA), the Max-Planck-Institute for Astrophysics (MPA), New Mexico State University, Ohio State University, University of Pittsburgh, University of Portsmouth, Princeton University, the United States Naval Observatory, and the University of Washington.  Funding for SDSS-III has been provided by the Alfred P. Sloan Foundation, the Participating Institutions, the National Science Foundation, and the U.S. Department of Energy Office of Science. The SDSS-III web site is http://www.sdss3.org/.  This work  made use of AstroPy \citep{astropy:2013, astropy:2018}, Matplotlib \citep{Hunter:2007}, NumPy \citep{numpy}, SciPy \citep{scipy}, and Pandas \citep{reback2020pandas}.

\appendix

\section{ICA Reconstructions}

Figure showing the ICA reconstruction of each of the 50 SDSS quasars in our sample that were observed by the VLA.  The \civ\ EW and blueshifts are derived from these reconstructions rather than from the original spectra.

%I used the following notebook to make stacks of spectra and their reconstructions, then concatenated them manually:  https://github.com/RichardsGroup/VLA2018b/blob/master/ICA_Reconstructions.ipynb
%On 3 March 2020, Amy R sent an email to TVM, GTR, PCH, and ABR with details regarding 49/50 of our reconstructions (not 1048+0032)
%On 25 May 2020, Amy R sent us the details for 1048+0032.
\begin{figure*}[th!]
    \epsscale{1.0}
    %\plotone{Figures/ICA_reconstructions}
    \plotone{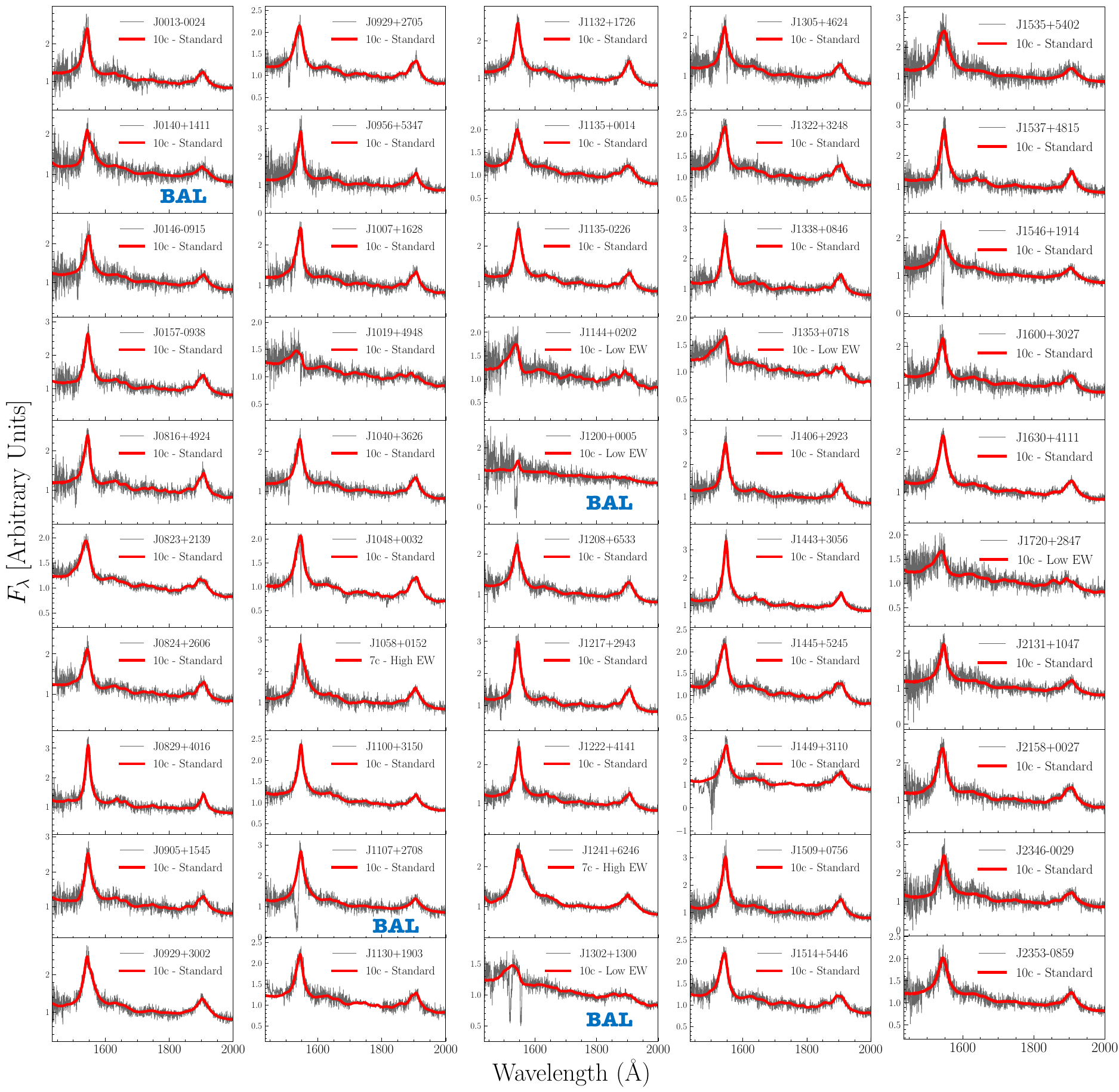}
    \caption{Spectra and the reconstructions from which \civ-emission parameters are measured.  The legend labels the abbreviated SDSSJ name of each spectrum's corresponding target and further specifies which of our three ICA reconstructions were applied to each target: the 10-component ``Standard" or ``Low" EW fit, or the 7-component ``High" EW; see \citet{Rankine+2020}.  BALQSOs are indicated based on classifications in \citet{Shen2011}.
    \label{fig:fig7new}}
\end{figure*}

%\bibliographystyle{mnras}
%\bibliography{references,stone}{}
%\bibliographystyle{aasjournal}

%\clearpage

\end{document}